\documentclass[useAMS,usenatbib]{mn2e}

\usepackage{epsf,rotating}
\usepackage{amsmath}
\usepackage{subfigure}

\newcommand{\kmsmpc}{\kms\;{\rm Mpc}^{-1}}
\newcommand{\lya}{Ly$\alpha$}
\newcommand{\HI}{\ion{H}{i}}
\newcommand{\OHI}{\Omega_{\rm HI}}

\newcommand{\hkpc}{h^{-1}{\rm kpc}}
\newcommand{\hmpc}{h^{-1}{\rm Mpc}}

\newcommand{\kms}{\;{\rm km}\,{\rm s}^{-1}}

\newcommand{\cmc}{\;{\rm cm}^{-3}}

\newcommand{\msolar}{\;{\rm M}_{\odot}}

\newcommand\cdunits{{\rm cm}^{-2}}

\newcommand{\gad}{{\sc Gadget-2}}

\newcommand{\ion}[2]{\hbox{#1\,{\sc #2}}}
\newcommand{\vw}{v_{\rm w}}

\title[HI in simulated galaxies]{The Neutral Hydrogen Content of Galaxies in Cosmological Hydrodynamic Simulations}

\author[Dav\'e et al.]{
\parbox[t]{\textwidth}{\vspace{-1cm}
Romeel Dav\'e$^{1,2,3,4}$, Neal Katz$^5$, Benjamin D. Oppenheimer$^6$, Juna A. Kollmeier$^7$, David H. Weinberg$^8$}
\\
\\$^1$ University of the Western Cape, Bellville, Cape Town 7535, South Africa
\\$^2$ South African Astronomical Observatories, Observatory, Cape Town 7925, South Africa
\\$^3$ African Institute for Mathematical Sciences, Muizenberg, Cape Town 7945, South Africa
\\$^4$ Astronomy Department, University of Arizona, Tucson, AZ 85721, USA
\\$^5$ Astronomy Department, University of Massachusetts, Amherst, MA 01003, USA
\\$^6$ Leiden Observatory, Leiden University, PO Box 9513, 2300 RA Leiden, Netherlands
\\$^7$ Observatories of the Carnegie Institute of Washington, Pasadena, CA 91101, USA
\\$^8$ Astronomy Department and CCAPP, Ohio State University, Columbus, OH 43210, USA
}

\begin{document}

\maketitle

 \begin{abstract}
We examine the global \HI\ properties of galaxies in quarter-billion
particle cosmological simulations using Gadget-2, focusing on how
galactic outflows impact \HI\ content.  We consider four outflow
models, including a new one (ezw) motivated by recent interstellar
medium simulations in which the wind speed and mass loading factor
scale as expected for momentum-driven outflows for larger galaxies
and energy-driven outflows for dwarfs ($\sigma<75\kms$).  To obtain
predicted \HI\ masses, we employ a simple but effective local
correction for particle self-shielding, and an observationally-constrained
transition from neutral to molecular hydrogen.  Our ezw simulation
produces an \HI\ mass function whose faint-end slope of $-1.3$
agrees well with observations from the ALFALFA survey; other models
agree less well.  Satellite galaxies have a bimodal distribution
in \HI\ fraction versus halo mass, with smaller satellites and/or
those in larger halos more often being \HI-deficient.  At a given
stellar mass, \HI\ content correlates with star formation rate and
inversely correlates with metallicity, as expected if driven by
stochasticity in the accretion rate.  To higher redshifts, massive
\HI\ galaxies disappear and the mass function steepens.  The global
cosmic \HI\ density conspires to remain fairly constant from $z\sim
5\rightarrow 0$, but the relative contribution from smaller galaxies
increases with redshift.
\end{abstract}


\section{Introduction}

The current paradigm for galaxy evolution rests on the tenet that
gas flows into and out of galaxies are primarily responsible for
governing galaxy growth~\citep[e.g.][and references
therein]{ker05,dek09,dav12}.  Such accretion and outflow processes
are often difficult to detect directly owing to the tenuous and
multi-phase nature of the gas, but they are expected to leave clear
imprints on the stellar and gaseous content of galaxies.  A key
region where such baryon cycling processes might be probed is in
the outskirts of star-forming galaxies, where reservoirs of neutral
hydrogen (\HI) are expected to hold the raw material that ultimately
fuels star formation.  This reservoir is thought to be continually
replenished by gravitationally-driven accretion from the highly
ionised intergalactic medium (IGM), possibly augmented by the
re-accretion of ejected gas.

Observations of the neutral hydrogen content of galaxies using 21cm
fine structure line emission are progressing rapidly, and promise
to improve further.  The \HI\ Parkes All-Sky Survey
\citep[HIPASS;][]{mey04} provided a comprehensive deep census of
the \HI\ content of low-redshift galaxies.  This was followed by
the Arecibo Fast Legacy Alfa Survey~\citep[ALFALFA;][]{gio05}, which
surveyed 7000~deg$^2$ for \HI\ 21cm emitting objects.  More recently,
the GALEX Arecibo SDSS Survey~\citep[GASS;][]{cat10} has obtained
high-fidelity \HI\ data on optically-selected galaxies, to provide
a more \HI-unbiased census.  Upgrades to the Jansky Very Large Array
(JVLA) promise further improvements.  Construction is underway for the MeerKAT
telescope, which will provide the ability to detect \HI\ in galaxies
out to unprecedented redshifts \citep[e.g. the Looking At the Distant
Universe with the MeerKAT Array -- LADUMA -- Survey;][]{hol11}, and
the Australian SKA Pathfinder (ASKAP), which will analogously perform
the Wallaby Survey.  Ultimately, the Square Kilometer Array (SKA)
will enable studies of \HI\ both nearby and out to higher redshifts
that far exceed today's capabilities.  Complementary information
is provided by studies of damped \lya\ absorbers (DLAs) that trace
\HI\ in and around galaxies in the spectra of background
quasars~\citep[e.g.][]{bat12,not12}.

It is, therefore, timely and crucial to develop a theoretical
framework for understanding the physics that governs the \HI\ content
of galaxies and its evolution.  Because galactic \HI\ reservoirs
are thought to represent a transient phase of baryons~\citep{pro09}
as they pass from the diffuse intergalactic medium~\citep[e.g.][]{dav10}
to higher density molecular gas, which eventually converts into
stars, it is essential that models for cosmic \HI\ be dynamical in
nature.  Cosmological hydrodynamic simulations provide one such
type of dynamical model for this purpose, allowing one to directly
track the physical state of gas as it flows through the \HI\
reservoir.  Particularly, the inclusion of galactic outflows in
simulations provides some interesting new twists on the accretion
paradigm, as enriched outflows that return back into galaxies may
provide a significant source of accretion~\citep[``wind
recycling";][]{opp10}.  These inflow, outflow, and recycling processes
are expected to manifest themselves in the \HI\ content of galaxies.

In this work, we examine the \HI\ content of galaxies in cosmological
hydrodynamic simulations with a variety of outflow models.  This
work follows on our earlier study in \citet{pop09}, but it uses a
significantly improved simulation suite and focuses on the physical
processes that govern the \HI\ content in addition to making
predictions for \HI\ observables and their evolution.  This work
is also comparable to that of \citet{duf12}, who used simulations
to examine both the atomic and molecular content of simulated
galaxies.  Compared to that work, our model explores a different
range of feedback parameters (including ones that match \HI\
observations substantially better), probes down to significantly
smaller systems (albeit in a smaller volume), and uses a slightly
different approach to compute the \HI\ content of galaxies.
Nonetheless, for overlapping models and mass ranges, our results
generally agree with those of \citet{duf12}.

The paper is organised as follows.  In \S\ref{sec:sims} we introduce
our simulations, and describe our method for calculating the atomic
and molecular gas content of our simulated galaxies.  In
\S\ref{sec:massfcn} we examine the \HI\ mass function and its
evolution, and we examine the complementary constraint of \HI\
richness versus stellar mass in \S\ref{sec:HIfrac}.  In \S\ref{sec:HIenv}
we study how \HI\ content is impacted by environment, and in
\S\ref{sec:HIdef} we show how it relates to galaxy metallicity and
star formation rate.  In \S\ref{sec:omegaHI} we make predictions
for the evolution of $\OHI$, and compare to observations out to
$z\sim 4$.  In \S\ref{sec:res} we discuss the robustness of
our \HI\ predictions to variations in resolution, molecular
gas prescription, and numerical method.  Finally, we summarise and
discuss our results in \S\ref{sec:summary}.

\section{Methods}\label{sec:sims}

\subsection{Simulations}

Our simulations are evolved with an extended version of the \gad~N-body
+ Smoothed Particle Hydrodynamic (SPH) code \citep{spr05}.  We
assume a $\Lambda$CDM cosmology~\citep{hin09}: $\Omega_{\rm M}=0.28$,
$\Omega_{\rm \Lambda}=0.72$, $h\equiv H_0/(100 \kmsmpc)=0.7$, a
primordial power spectrum index $n=0.96$, an amplitude of the mass
fluctuations scaled to $\sigma_8=0.82$, and $\Omega_b=0.046$.  We
call this cosmology our r-series, where our general naming convention
is r[{\it boxsize}]n[{\it particles/side}][{\it wind model}].  Our
primary simulations use a cubic volume of $32\hmpc$ on a side with
$512^3$ dark matter and $512^3$ gas particles, and a softening
length of $\epsilon=1.25\hkpc$ (comoving, Plummer equivalent).  The
gas particle mass is $4.5\times 10^6 M_\odot$, and the dark matter
particles masses are approximately five times larger.  We can thus
reliably resolve galaxies down to stellar masses of $M_{\rm
*,lim}=1.4\times 10^8 M_\odot$ (see discussion below).

Our version of \gad\ includes cooling processes using the primordial
abundances as described by \citet{kat96}, with additional cooling
from metal lines assuming photo-ionisation equilibrium from
\citet{wie09}.  Star formation is modelled using a subgrid recipe
introduced by \citet{spr03a} where a gas particle above a density
threshold of $n_{\rm H}=0.13 \cmc$ is modelled as a fraction of
cold clouds embedded in a warm ionised medium following \citet{mck77}.
Star formation (SF) follows a Schmidt law \citep{sch59} where
the SF rate is proportional to $n_{\rm H}^{1.5}$, with the
SF timescale scaled to match the $z=0$ Kennicutt relation \citep{ken98}.
We use a \citet{cha03} initial mass function (IMF) throughout.  We
account for metal enrichment from Type II supernovae (SNe), Type
Ia SNe, and AGB stars, and we track four elements (C,O,Si,Fe)
individually, as described by \citet{opp08}.  We note that the
entropy-conserving SPH algorithm in \gad\ is known to have some
deficiencies in properly modelling hydrodynamical instabilities; in
\S\ref{sec:res} we discuss this further and argue that this should 
not have a large impact on our results.

Galactic outflows are implemented using a Monte Carlo approach
analogous to star formation.  Outflows are directly tied to the
SFR, using the relation $\dot M_{\rm wind}= \eta \times$SFR, where
$\eta$ is the outflow mass loading factor.  The probability for a
gas particle to spawn a star particle is calculated from the subgrid
model described above, and the probability to be launched in a wind
is $\eta$ times the star formation probability.  If the particle
is selected to be launched, it is given a velocity boost of $v_w$
in the direction of {\bf v}$\times${\bf a}, where {\bf v} and {\bf
a} are the particle's instantaneous velocity and acceleration,
respectively.  Once a gas particle is launched, its hydrodynamic
(not gravitational) forces are turned off until either
$1.95\times10^{10}/(\vw (\kms))$ years have passed or, as more often
occurs, the gas particle has reached a density that is 10\% of the
SF critical density (i.e., 0.013~cm$^{-3}$).  This attempts to mock
up chimneys generated by outflows that would allow a relatively
unfettered escape from the galactic ISM, a process not properly
captured by the spherically-averaging SPH algorithm at $\sim$kpc
resolution.  It also yields results that are less sensitive to
numerical resolution~\citep{spr03b} than models that do not turn
off hydrodynamic forces.  For a further discussion of hydrodynamic
decoupling, see \citet{dal08}.

Choices of the parameters $\eta$ and $v_w$ define the ``wind model".
For this paper we make use of the following four wind models:\\
(i) {\bf No winds (nw)}, where we do not include outflows (i.e. $\eta=0$).
This model fails to match a wide range of observables
\citep[e.g.][]{dav11a,dav11b}, but is included to establish a baseline
for the overall impact of winds.\\
(ii) {\bf Constant winds (cw),} where $\eta=2$ and $\vw=680 \kms$ for all
galaxies.  This model is similar to the wind model employed by \citet{duf12}.\\
(iii) {\bf Momentum-conserving winds (vzw),} where the wind speed and
the mass loading factor depend on the galaxy velocity dispersion
$\sigma$~\citep{mur05}, using the relations~\citep[see][]{opp08}.  
\begin{eqnarray}
  \vw &=& 3\sigma \sqrt{f_L-1}, \label{eqn: windspeed} \\
  \eta &=& \frac{\sigma_0}{\sigma} \label{eqn: massload},
\end{eqnarray} 
where $f_L=[1.05,2]$ is the luminosity factor in units of the
galactic Eddington luminosity (i.e. the critical luminosity necessary
to expel gas from the galaxy potential), and $\sigma_0=150$~km/s
is the normalisation of the mass loading factor.  Choices for the
former are taken from observations~\citep{rup05}, while the latter
is broadly constrained to match high-redshift IGM enrichment~\citep{opp08}.  
Galaxies are identified ``on-the-fly'' during the simulation using a
Friends-of-Friends (FOF) algorithm applied to the gas, star, and dark matter
particles.  We choose a smaller than usual linking length to pick out just
the galaxies. This linking length evolves with redshift and in terms of the
mean interparticle separation of all particles is
\begin{equation}\label{eqn:linkinglength}
0.06 \Bigl(\frac{H(z)}{H_0}\Bigr)^{1/3}.
\end{equation}
We estimate the velocity dispersions necessary for our wind scaling laws 
from the total FOF galaxy mass $M_{\rm gal}$\footnote{We have actually
been using this algorithm, which slightly differs from \citet{opp08}, in all of
our recent simulations starting with \citet{opp10}.} using the relation
\begin{equation}\label{eqn:sigma}
\sigma = 200 \Bigl(\frac{M_{\rm gal}}{5\times 10^{12} h^{-1} M_\odot} 
\frac{H(z)}{H_0} \Bigr)^{1/3} \;\; {\rm km/s},
\end{equation}
which we have empirically determined gives an accurate measure of the velocity
dispersion in our simulations.

(iv) {\bf Hybrid energy/momentum-driven winds (ezw),} which employs
the vzw scalings for galaxies with $\sigma>75\,\kms$, then switches
over to a steeper dependence of $\eta\propto \sigma^{-2}$ in
$\sigma<75\,\kms$ systems.  The wind speed still scales proportionally
to $\sigma$ as in the vzw model.  This model roughly captures the
behaviour in recent analytic and hydrodynamic models of outflows
from interstellar media by \citet{mur10} and \citet{hop12},
respectively.  The basic idea is that in dwarf galaxies the energy
from supernovae plays a dominant role in driving outflows, while
in larger systems the momentum flux from young stars and/or supernovae
is the dominant driver.  As a result, the outflow scalings switch
from momentum-driven at high masses to energy-driven at low masses.
We make this transition abruptly at $\sigma=75\,\kms$ guided by the
analytic models of \citet{mur05,mur10} and the high-resolution ISM
simulations of \citet{hop12}, although it should perhaps be more
gradual (note that $\eta$ itself is continuous across this boundary).
In any case, this model captures the gist of the most up-to-date
small-scale outflow models.  As we will see, the ezw outflow model
fares somewhat better compared to observations than vzw, which in
turn fares much better than cw or nw.

Additionally, in our fiducial ezw simulation, we employ a heuristic
prescription to quench star formation in massive galaxies.  This
is not a physical model, but simply a tuned parametrisation to limit
star formation in massive systems to reproduce the observed exponential
high-mass cutoff in the stellar mass function and to make the
simulation faster to run.  In this prescription, we quench star
formation in an {\it entire galaxy} according to a probability,
$P_{Q}$, given by the equation \begin{equation}\label{eqn:quench}
P_{Q} = 1 - \frac{1}{2}{\rm erfc}\Bigl[\frac{\log(\sigma) -
\log(\sigma_{\rm Qmed})}{\log(\sigma_{\rm Qspr})} \Bigr], \end{equation}
where the median $\sigma$ at which a galaxy has a 50\% chance of
being quenched is $\sigma_{\rm Qmed} = 110 \kms$, corresponding to
$M_{\rm halo}=10^{12.1} \msolar$ at $z=0$, and the spread in $\sigma$
is $\sigma_{\rm Qspr} = 32 \kms$.  We also require $\sigma >75 \kms$
to fully suppress the already low probability of lower mass galaxies
being quenched.  We have found that these parameter choices nicely
reproduce the high-mass end of the stellar mass function, as we
will show in \S\ref{sec:gsmf}.

Every time we identify galaxies using our FOF group finder, which
we do to calculate $\sigma$ for the vzw and ezw wind models
every 10~Myr, we compute this quenching
probability for each identified galaxy.  If the galaxy is ``quenched",
then each time a gas particle would have formed a star over the
time interval until the next time we identify galaxies using our
FOF group finder, it is instead heated to 50 times the virial
temperature, $T_{\rm vir}$, where
\begin{equation}\label{eqn:Tvir}
T_{\rm vir} = 5\times 10^6 \Bigl(\frac{\sigma}{200 \kms}\Bigr)^{2} {\rm K.}
\end{equation}
The motivation for heating the gas to such
extreme temperatures is primarily to prevent it from re-accreting at later times
and thus requiring multiple ejections; the total energy input is thus less for
higher quenching temperatures.  
The median temperature to which the ISM is heated by
quenching is $10^{8.1}$ K and arises from a median halo mass of
$10^{12.2} \msolar$.  
The energy input from quenching
averages to $10^{40.5} {\rm erg s}^{-1} {\rm Mpc}^{-3}$ between
$z=0.75-2.5$, corresponding to the peak of AGN activity, and the
integrated energy input until $z=0$ is $8\times 10^{14}$ ergs/g, which
equals $9\times10^{-7}$ of the rest mass energy of all baryons.
Considering that 6.5\% of baryons are in stars in this simulation at
$z=0$, and using the assumption that supermassive black holes (SMBHs)
have $10^{-3}$ of the mass in stars, the quenching energy corresponds
to 1.4\% of the rest-mass energy of SMBHs, which is comparable to the
energy imparted from AGN feedback in cosmological
simulations that self-consistently track black hole growth
and feedback \citep{dim05,boo09}.

In this paper we are mostly concerned with galaxies around $L^*$
and below, which are mostly unquenched.  We emphasise that this
quenching prescription is not intended to be a realistic physical
model, and is actually in large part a computational convenience,
as removing gas from the largest galaxies substantially speeds up
the simulation at low redshifts and thus allows us to run our large
simulations to $z=0$ within practical time frames.  We refer the
reader to \citet{duf12} for a discussion of \HI\ in larger-volume
simulations that attempt to quench massive galaxies based on directly
tracking black hole growth and feedback.

Finally, our no-wind simulation employs a volume of $16\hmpc$ with
$2\times 256^3$ particles.  This results in the same mass and spatial
resolution as the wind simulations, but in a volume that is $8$
times smaller.  As we mostly use the no-wind simulation to qualitatively
show the impact of winds at small masses, the reduced volume for
the simulation will not significantly affect our conclusions.

\subsection{Computing the \HI\ content}

We use SKID\footnote{http://www-hpcc.astro.washington.edu/tools/skid.html}
(Spline Kernel Interpolative Denmax) to identify galaxies as bound
groups of star-forming gas and stars~\citep{ker05,opp10}.  Our
galaxy stellar mass limit is set to be $\ge 64$ star
particles~\citep{fin06}, resulting in a minimum resolved mass of
$M_{\rm *,min}=1.4\times 10^8 M_\odot$.  We will only consider
galaxies with stellar masses $M_*\geq M_{\rm *,min}$ in our analysis,
regardless of their \HI\ content.  Our resolution convergence tests
in \S\ref{sec:res} show that the \HI\ properties are reasonably
well converged even at this stellar mass threshold.

We identify dark matter halos via a spherical overdensity
algorithm~\citep{ker05} out to a virial overdensity given in
\citet[see their eq.~1]{dav10}, centred on each galaxy.  We separate
galaxies into central and satellite galaxies by associating each
galaxy with a halo; if a galaxy's centre lies within the virial
radius of a larger (by stellar mass) galaxy, we consider it to be
a satellite of that galaxy, and the halos of those two galaxies are
merged.  Note that for galaxies near the edge of a larger halo,
this can result in a halo that has ``bumps" along its (mostly
spherical) surface.

\begin{figure}
\vskip -0.3in
\setlength{\epsfxsize}{0.55\textwidth}
\centerline{\epsfbox{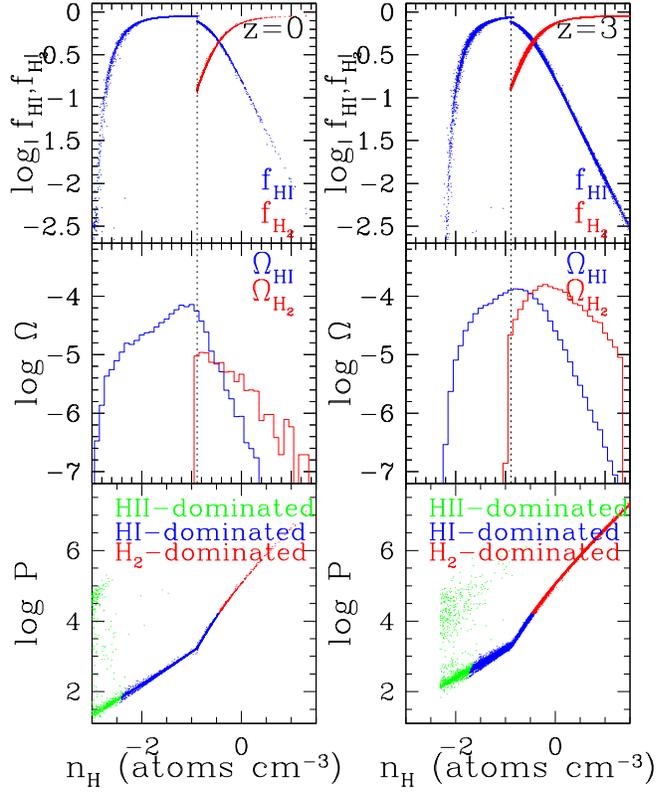}}
\vskip -0.1in
\caption{Gas particle properties as a function of number density
$n_H$ in our r48n384vzw simulation.  Left panels show $z=0$, right
panels show $z=3$.  {\it Top:} \HI\ fraction (blue) and H$_2$
fraction (red) based on the calculation described in the text. We show a
random sampling of 1\% of all the particles.  {\it Middle:}
Total mass density $\Omega$ per unit $\log n_H$ in \HI\ (blue) and
$H_2$ (red).  {\it Bottom:} Pressure-density relation.  Green points
show unshielded particles (either below the pressure threshold or
above the density threshold), blue points show \HI-dominated
gas, and red points show molecular-dominated gas.
}
\label{fig:fneut}
\end{figure} 

To compute the \HI\ content of galaxies, we need to isolate the gas
that is in the (predominantly) neutral phase.  This involves defining
two ``boundaries":  the division between gas that is exposed to the
full metagalactic ionising flux and gas that is self-shielded, and
the division between atomic and molecular gas.

To calculate the self-shielding, we first compute the neutral
hydrogen component of each gas particle under the assumption that
it is {\it not} self-shielded.  We follow \citet{pop09} who employed
a simple hydrogen ionisation balance calculation that yielded the
following formula to determine the neutral fraction:
\begin{equation}
f_{\rm HI} = \frac{2C+1-\sqrt{(2C+1)^2-4C^2}}{2C}
\end{equation}
with
\begin{equation}
C = \frac{n \beta(T)}{\Gamma_{\rm HI}}
\end{equation}
where $n$ is the Hydrogen number density, $T$ is the gas temperature,
$\Gamma_{\rm HI}$ is the \ion{H}{i} photo-ionisation rate, and the
recombination rate coefficient $\beta$ is well-fit by 
the function~\citep{ver96}
\begin{equation}
\beta(T) = a\Bigl[ \sqrt(T/T_0)(1+\sqrt(T/T_0))^{1-b}(1+\sqrt(T/T_1))^{1+b}\Bigr]^{-1}.
\end{equation}
For \HI, the fitting parameters are $a=7.982\times 10^{-11}$~cm$^3$s$^{-1}$, 
$b=0.7480$, $T_0=3.148$~K, and $T_1=7.036\times 10^5$~K.  We take
$\Gamma_{\rm HI}$ from the \citet{haa01} ionising background, whose amplitude
is adjusted to match the observed mean flux decrement in the
\lya\ forest in these simulations~\citep[see][]{dav10}; we describe this
further below.

With the optically-thin neutral fraction computed for each gas
particle, we employ a simple particle-by-particle post-processing
correction for self-shielding of the \HI.  We resort to this approach
because it is computationally prohibitive to do the full radiative
line transfer on these simulations; instead, we will calibrate our
approach to radiative transfer simulations by \citet{fau10}.

We begin by assuming that each particle has a density profile given
by the SPH spline kernel $W(r)$~\citep[see][for definition]{spr05}.
We compute the radial column density profile as follows:
\begin{equation}
N_{\rm HI}(r) = \frac{0.76 f_{\rm HI} \rho_g}{m_p}\int^{h}_r W(r') dr',
\end{equation}
where $\rho_g$ is the SPH density of the gas particle, $m_p$ is the proton
mass, and $h$ is the particle's smoothing length.  We then determine the
radius $R$ where $N_{\rm HI}(R)=N_{\rm HI,lim}$, with 
$N_{\rm HI,lim}=10^{17.2}\cdunits$ 
being where the particle becomes optically thick ($\tau=1$) to continuum
photons at the Lyman limit.  If no such radius $R$ exists,
then the particle is optically thin.  If $R$ exists, then we compute
the unshielded and shielded mass fraction using $4\pi\int W(r)r^2dr$
integrated from $h\rightarrow R$ and $R\rightarrow 0$, respectively.
The neutral fraction of the particle is then the mass-weighted mean
neutral fraction, assuming that the portion from $R\rightarrow 0$ is
90\% neutral (since some ionised gas is seen to exist even within dense
regions).  By this procedure, particles in dense regions ``auto-shield"
themselves from the ambient flux, though the effects of shielding from
nearby gas are neglected.

This computation relies on knowing the metagalactic photo-ionising
flux, since this determines the neutral fraction in the optically-thin
regime. Our simulations assume a \citet{haa01} ionising background,
but more detailed constraints can be placed on the amplitude of the
\HI\ photo-ionisation rate by using the measured mean flux decrement
in the Lyman alpha forest. In particular, at each redshift for each
simulation, we determine a ``flux factor", which is the value by
which we must multiply the strength of the \citet{haa01} background
to achieve consistency with the observed flux decrement.  For the
observed flux decrement, we employ at $z<2$ the determination by
\citet{kir07}, while for $z\geq2$ we take the values from \citet{bec12}.
As described more fully in \citet{dav10}, we extract 1000 spectra
and iteratively adjust the flux factor until the mean flux is within
1\% of the observed value.  The resulting flux factors for our ezw
simulation at $z=0,1,2,3,4,5$ are 1.72, 2.03, 1.66, 0.92, 1.13, and
1.70, respectively.  For our vzw simulation, the $z=0$ flux factor
is 1.63 and $z=3$ is 0.94, while for cw, it is 1.41 at $z=0$ (we
do not consider these models at other redshifts).

Next, we separate the molecular component from the atomic \HI.  To
do so, we employ the observed ISM pressure relation from The \HI\
Nearby Galaxy Survey~\citep[THINGS;][]{ler08}; specifically, we use
their ``combined spiral subsample" fit, which gives the ratio of
molecular to atomic gas as \begin{equation} R_{\rm mol} = (P/P_0)^\alpha,
\end{equation} where $P_0=1.7\times 10^4$~cm$^{-3}$K and $\alpha=0.8$.
We compute the gas pressure based on the density and two-phase
medium temperature of each star-forming gas particle~\citep[our
prescription follows][]{spr03a}.  Note that we only compute molecular
fractions for star-forming gas particles, which in our simulations
is gas with $n_{\rm H}>0.13$~cm$^{-3}$.  Gas that is not star-forming
is assumed to have zero molecular content, and in any case the
formula above would yield a very small molecular fraction.  This
prescription follows that employed by \citet{duf12}, but differs
from \citet{pop09} who used a fixed pressure threshold of 810~cm$^{-3}$K
to separate atomic from molecular gas.

Recent theoretical work by \citet{kru11} has also provided a
prescription for separating molecular from atomic gas, which includes
a metallicity dependence.  We describe this method and examine its
impact in \S\ref{sec:res}.  To encapsulate the results, this
theoretically-based prescription has little effect at $z=0$, but
at $z=3$ the lower metallicity causes somewhat less molecular gas
to form, and hence the \HI\ content is slightly higher.

Figure~\ref{fig:fneut} illustrates the resulting gas particle atomic
and molecular fractions.  The top left panel shows the \HI\ and
H$_2$ mass fractions in particles from our momentum-driven wind
scaling (vzw) simulations at $z=0$.  In the optically thin
regime~\citep[mostly not depicted here; see][for a full phase space
diagram]{dav10}, $f_{\rm HI}$ scales linearly with $n_{\rm H}$.
Then there is a sharp upturn at $n_{\rm H}\sim 2\times 10^{-3}$cm$^{-3}$,
when auto-shielding becomes important.

At high redshifts, the stronger ionising background causes
auto-shielding to set in at a higher density.  A simple scaling
shows that $N_{\rm HI,lim} \propto f_{\rm HI} n_{\rm H} l$, where
the length $l \propto n_{\rm H}^{-1/3}$ for a spherical cloud (or
SPH particle), and $f_{\rm HI}\propto n_{\rm H}/\Gamma_{\rm HI}$;
hence $N_{\rm HI,lim} \propto n_{\rm H}^{5/3}/\Gamma_{\rm HI}$.
Therefore, a factor of $10$ increase in $\Gamma_{\rm HI}$, which
occurs between $z=0\rightarrow 3$, will lead to a factor of $4$
increase in the $n_{\rm H}$ where self-shielding becomes important,
because $n_{\rm H}\propto \Gamma_{\rm HI}^{3/5}$ for constant $N_{\rm
HI,lim}$.  Figure~\ref{fig:fneut} (top right panel) shows that
auto-shielding at $z=3$ is effective above $n_H\sim 10^{-2}$cm$^{-3}$
as expected for the stronger $z=3$ background UV field, which also
matches the $z=3$ radiative transfer simulations of \citet[see their
Figure~3]{fau10}.  Indeed the overall shape of $f_{\rm HI}(n_H)$
is actually quite similar to theirs, although there is less scatter
at a given $n_H$ in our prescription owing to the fact that we do
not consider shielding from neighbouring particles.  This shows
that our physically-motivated choice of auto-shielding with $N_{\rm
HI,lim}$ at the Lyman limit is a reasonable approximation to much
more detailed radiative transfer models.

Moving to higher densities, eventually we reach the star-forming
density threshold (the vertical dashed line), above which the
molecular fraction becomes nonzero.  At $n_H\ga 0.5$ cm$^{-3}$,
corresponding to a pressure close to $P_0$, the gas becomes
molecular-dominated.  The pressure relation is shown in the bottom
panels of Figure~\ref{fig:fneut}, with the particles colour-coded
by their dominant phase of hydrogen.  The THINGS pressure threshold
is much higher than the $P/k=810$~cm$^{-3}$K assumed in \citet{pop09}
and, moreover, our prescription produces a more gradual transition
between atomic and molecular-dominated gas.  A change in slope at
$n_H = 0.13$ cm$^{-3}$ occurs because this is the density above
which we allow stars to form and the gas particles become two-phase.

In this paper we are mostly interested in the regime $10^{-2.5}\la
n_H\la 0.5$~cm$^{-3}$ where the majority of cosmic \HI\ resides (at
$z=0$), above which gas becomes mostly molecular and below which
it is mostly ionised.  This is shown in the middle panels of
Figure~\ref{fig:fneut}, where we plot the total mass density $\Omega$
in \HI\ (blue) and $H_2$ (red) per unit interval of $\log{n_H}$;
the peak is around $n_H\sim 10^{-1}$~cm$^{-3}$.  This \HI-dominant
density regime is reasonably well resolved in our simulations, and
hence the predicted \HI\ content is expected to be robust, despite
it being a transitory phase.  In contrast, the transition from
molecular gas to stars typically occurs at densities well above
what we can resolve directly, and hence the molecular content may
not be quite as robustly predicted.  Still, our star formation 
prescription is consistent with the~\citet{ken98} relation, which
is well-established on the $\sim$kpc scale resolution of our
simulations.  Our predictions for the evolution from $z\sim 2\rightarrow
0$ of the total star-forming gas
in our momentum-driven wind (vzw) simulation are in good
agreement with observations from the Plateau de Bure High-$z$ Blue
Sequence Survey~\citep[PHIBSS;][see their Figure~13]{tac12}; other
wind models fare less well.  Finally, we note that the stellar mass
growth rate is fairly robust because it is not a transitory phase
but an end state of accreted gas, and is typically limited by the
gas supply rate modulated by outflows~\citep[e.g][]{fin08,bou10,dav12}.

We note that, by the above prescription to divide molecular from
atomic gas, 55\% of the gas at $z=0$ that is above our star 
formation density threshold of $n_H=0.13$~cm$^{-3}$ is actually 
neutral (it is much lower at high redshift).  This gas could in principle
form stars in our simulation,
which is contrary to the idea that stars only form from molecular
gas.  However, the amount of star formation actually occurring in
this neutral gas is very small, only about 3\%.  This is because
the star formation rate scales as $\rho^{1.5}$, which means that
virtually all of the star formation occurs in the denser gas
that is almost fully molecular.

With each gas particle's \HI\ content determined, we must now
associate the gas particles with galaxies.  The information from
SKID is not sufficient, because SKID only includes star-forming gas
and stars in a galaxy, while a significant amount of the \HI\ resides
in an extended region around the galaxy, beyond the actively
star-forming region.  Thus to account for extended \HI, we add up
all the \HI\ mass in a sphere around each galaxy that extends to
the outermost radius as defined by SKID, i.e. the radius of the
farthest SKID particle associated with that galaxy.  The outermost
radius is typically many times the half-mass radius.  While it may
seem that this still may not fully account for an extended \HI\
disk, in practise the low threshold density for star formation in
our simulations means that this choice still encompasses the vast
majority of the \HI.  We tried extending this to 1.5 times the
outermost radius, and the total \HI\ mass in the volume increased
by only 4\%; furthermore, one will start to increasingly ``double
count" gas that may be between nearby galaxies as being part of
both galaxies.  Modestly reducing this radius also has minimal
effect, so our \HI\ masses are not very sensitive to this choice.

\subsection{The Stellar Mass Function}\label{sec:gsmf}

\begin{figure}
\vskip -0.2in
\setlength{\epsfxsize}{0.6\textwidth}
\centerline{\epsfbox{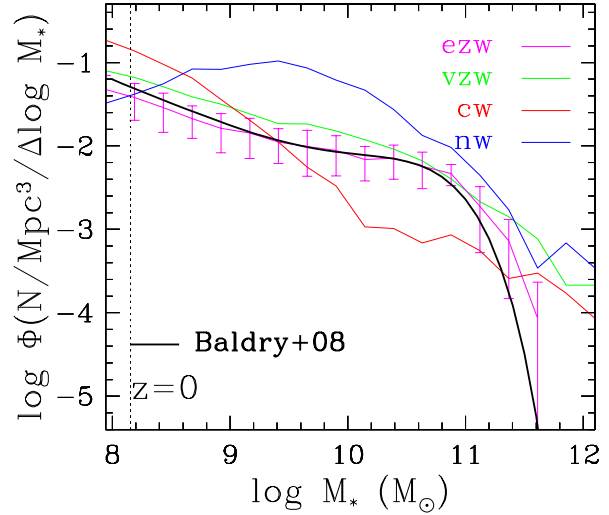}}
\vskip -2.5in
\caption{Galaxy stellar mass functions at $z=0$ in our four
wind models: ezw (green), vzw (cyan), cw (red), and nw (blue).
The vertical dotted line is the stellar mass resolution limit.
The thick solid line shows a fit to the observed GSMF from \citet{bal08}.
The ezw outflow model, which includes a heuristic prescription for quenching
star formation at high masses, matches the observed GSMF to within 
uncertainties.
}
\label{fig:gsmf}
\end{figure} 

As an initial baseline statistic to compare our four wind models,
we show in Figure~\ref{fig:gsmf} their galaxy stellar mass functions
(GSMFs) at $z=0$.  We compare these to observations from \citet{bal08}
using SDSS.

The ezw model provides a strikingly good fit to the observed GSMF.
At the massive end, this is a direct consequence of tuning the
quenching prescription as described previously.  In contrast, the
low-mass end is unaffected by our quenching prescription, and instead
reflects the effect of the hybrid energy-momentum driven winds.
There is a clear upturn in the GSMF at $M_*\la 10^{9.5}M_\odot$,
which is also seen in the data.  As explained in \citet{opp10}, in
our simulations this arises because above this mass, wind recycling
becomes increasingly important, and provides extra fuel to higher
mass galaxies.  Below this mass, the typical recycling time becomes
longer than a Hubble time, and the slope begins to steepen towards
the dark matter halo mass function's slope.

The vzw model provides not quite as good a fit to the low mass end,
similar to what was seen in a lower-resolution version of the same
model in \citet{dav11a}.  The differences at the massive end
effectively show the impact of the quenching model, since our ezw
simulation uses it while our vzw simulation does not, while the two
outflow prescriptions themselves are identical in this mass regime.
Quenching has a substantial effect on the GSMF, but we will show
that it has a minimal effect on the \HI\ mass function.  Meanwhile,
the constant wind model produces a very steep low mass end slope
that looks nothing like the data, while the no wind model overproduces
stars at virtually all masses because it strongly overcools baryons.
These results again follow those obtained using lower-resolution
simulations of the same wind models presented in \citet{opp10} and
\citet{dav11a}, and more discussions of the GSMF can be found there.

This ezw simulation is the first hydrodynamic simulation that we
are aware of to yield agreement with the observed GSMF to within
statistical uncertainties over the entire mass range probed.  This
does not imply that this model is fully correct, as it could still
be that the star formation histories in this model are incorrect,
and other constraints may not be as well matched.  As a case in
point, we will show in \S\ref{sec:HIdef} that the mass-metallicity
relation in this model looks too steep compared to observations,
and that the vzw model provides a better match.  Furthermore, the
success of this model at high masses owes to including an ad hoc
and highly tuned prescription for quenching star formation that is
not a direct implementation of a physical quenching mechanism.  We
leave a full comparison of the ezw model to a wide range of observables
for future work, and focus here on the \HI\ properties of galaxies
in our four wind simulations.

\section{The HI Mass Function}\label{sec:massfcn}

\begin{figure}
\vskip -0.2in
\setlength{\epsfxsize}{0.6\textwidth}
\centerline{\epsfbox{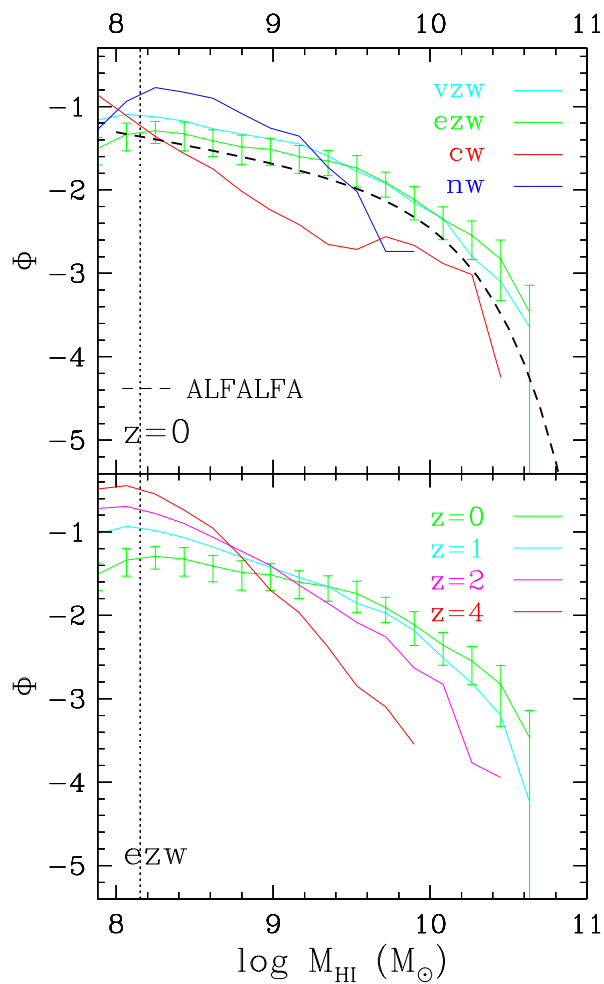}}
\vskip -0.2in
\caption{{\it Top:} A comparison of the HIMF at $z=0$ in our four wind
simulations -- ezw (green), vzw (cyan), cw (red), and nw (blue).  
The vertical dotted line is the stellar
mass resolution limit of $1.4\times 10^8 M_\odot$; 
this is roughly the \HI\ mass resolution limit
as well since $M_*$ and $M_{HI}$ are comparable at the smallest masses
(see Figure~\ref{fig:HIfrac}).  The dashed line shows the HIMF from the
$\alpha.40$ sample of the ALFALFA survey \citep{hay11} for comparison.
The ezw and vzw model provide reasonable fits to the data.  
{\it Bottom:} Evolution of the HI mass function from $z=4\rightarrow 0$ in the 
ezw model.  The low mass end slope increases substantially at high redshifts.
}
\label{fig:massfcn}
\end{figure} 

The most basic counting statistic for characterising the \HI\ content
of galaxies is the \HI\ mass function (HIMF).  Improving 21cm
observations have enabled the HIMF to be probed down to \HI\ masses
approaching $10^7 M_\odot$~\citep[e.g.][]{zwa05,hay11}.  However,
the redshift evolution of the HIMF remains poorly characterised,
awaiting the next generation of facilities.  Here we compare our
simulated HIMFs with observations, to understand what constraints
can be placed on wind models and how the HIMF is expected to evolve.

Figure~\ref{fig:massfcn}, top panel, shows the HIMF for our four
wind models.  The vertical dotted line shows the {\it stellar} mass
resolution limit of our fiducial r32n512ezw simulation.  We have
no formal \HI\ mass resolution limit, but we will show in \S\ref{sec:res}
that this is a reasonable choice for ensuring resolution convergence 
of the HIMF as well.  Error bars (shown for ezw)
depict cosmic variance as computed from the error of the mean HIMF
in each of eight octants of the simulation volume.

For comparison, we show the observational determination of the HIMF
from \citet{hay11} (the first 40\% of the ALFALFA Survey; dashed
line).  We note that the ALFALFA observations select by HI mass,
while we are effectively selecting by stellar mass; however, the
sensitivity of ALFALFA is well below what we can resolve, and hence
unless there is a dominant population of small HI-free objects
(which we will show later does not occur in our models), this
comparison should be robust.

Concentrating on the low mass end of the HIMF, we see that both the
ezw and vzw models do a good job of matching the observed low mass
end slope of $\alpha\approx -1.34\pm0.02$~\citep[for the ``whole
$\alpha.40$" sample, from Table~6 of][]{hay11}.  In detail, the
predicted best-fit Schechter function slope for the ezw model is
$\alpha=-1.31$, while for the vzw model it is $\alpha=-1.45$ and,
therefore, the ezw model provides a slightly better fit.  The massive
end is relatively unaffected by quenching, as seen from the minor
differences between the ezw model that includes quenching and the
vzw that does not include it.

Particularly remarkable is that the amplitudes of the HIMF in these
two models are reasonably close to that observed.  They are slightly
above the observations at all masses, particularly at higher masses.
We could in principle tune our values of $N_{\rm HI,lim}$ and/or
shrink the radial extent to which we associate \HI\ with a galaxy
to match the data better, but we
prefer to use well-motivated values for these quantities.  We have
tried several other reasonable values for these quantities,
and the net effect is generally to scale the HIMF in amplitude
without changing the shape significantly.

The agreement with the low mass end slope of the HIMF, while
simultaneously matching the low mass end of the galactic stellar
mass function, is a stringent constraint that almost all galaxy
formation models have difficulty matching~\citep{mo05,lu12,lu13}.
Fundamentally, in most models this arises because the low mass end
slope of the dark matter halo mass function is quite steep, which
is exacerbated by low-mass galaxies being more \HI-rich.  In our
simulations, the stronger outflows from low-mass galaxies strongly
suppress the overall baryon content of galaxies.  Even more critical
is the impact of wind recycling.  As described in \citet{opp10},
wind recycling is preferentially stronger in high-mass galaxies
because the denser surrounding gas slows outflows more effectively
via ram pressure.  This results in a higher fraction of ejected gas
being recycled back into higher-mass galaxies, thereby yielding
more star formation in more massive systems.  \citet{opp10} showed
that this flattens the low mass end slope of the stellar mass
function (see Figure~\ref{fig:gsmf}), and here we see that this
effect is also important for the low mass end of the \HI\ mass
function.  We will discuss further the differences between previous
semi-analytic galaxy formation model results and our results in
\S\ref{sec:summary}.

Both the constant wind (cw) and no wind (nw) models provide
significantly poorer fits to the observations than ezw or vzw.
Without winds, there is a surplus of low-$M_{\rm HI}$ galaxies, and
a deficit of high-$M_{\rm HI}$ systems.  This is characteristic of
the HIMF in many semi-analytic galaxy formation models~\citep{lu12}.
The former discrepancy arises because there are no outflows to eject
material from low-mass galaxies, and hence the baryon fraction in
these systems is very large~\citep{dav09}, in disagreement with
observations that show a small baryon content in
dwarfs~\citep[e.g.][]{mcg10}.  The latter arises because, without
winds, gas becomes very dense in high-mass galaxies, which means
that most of the gas is in molecular form, and much of it has been
converted into stars.  As discussed in \citet{dav11a}, the stellar
content of galaxies in the no-wind model grossly exceeds that
observed, and Figure~\ref{fig:gsmf} showed that it overproduces the
number density of galaxies at all the stellar masses probed.

For the constant wind case, the low mass end slope is quite steep
($\alpha=-1.64$).  Moreover, there is a characteristic ``bump" in
the mass function at $M_{\rm HI}\sim 10^{10}M_\odot$.  This bump
is reminiscent of a similar bump in the stellar mass function in
this model~\citep{dav11a}, which arises because wind recycling
rapidly becomes important around this mass scale as the (constant
velocity) winds are no longer able to escape the galaxy halo's
potential well.  Such a feature, which is generic to wind models
that assume a constant outflow speed, is not observed in the HIMF
(nor in the stellar mass function).

\citet{duf12} employed an outflow model that is similar to this
constant wind case.  They find that the low mass end slope is fairly
flat down to $M_{\rm HI}\approx 10^9 M_\odot$, and was too shallow
compared to data.  In a comparable range, our HIMF also appears
fairly flat -- formally, the best-fit low mass end slope ignoring
points below that mass is $\alpha=-1.12$, although a Schechter
function is a poor descriptor.  Our higher-resolution simulation
probes further down the mass function, which enables us to see the
steep low mass end.  Overall, our results for this wind model agree
well with \citet{duf12} in the overlapping mass range, and both
show that the constant wind model fails to match the low mass end
of the HIMF.

The bottom panel shows the redshift evolution of the HIMF out to
z=4, focusing on our ezw simulation.  The low mass end slope becomes
progressively steeper with redshift, mimicking the behaviour seen
in the stellar mass function~\citep{dav11a}.  The low mass end
slopes at $z=1,2,3,4,5$ are $-1.54,-1.79,-1.82,-1.99,-2.11$,
respectively ($z=3,5$ are not shown).  This arises because wind
recycling becomes increasingly effective at lower redshifts~\citep{opp10},
since the recycling time is roughly constant at $\sim
1-3$~Gyr~\citep{opp08} for $M_*\sim 10^{10}M_\odot$ galaxies, which
is a small fraction of the Hubble time today but comparable to the
Hubble time in the early Universe.  Hence the HIMF steepens rapidly
out to $z\sim 2$, and then the steepening becomes more gradual,
since at early times the effect of wind recycling, which is responsible
for flattening the HIMF, is reduced.  Meanwhile at the massive end,
there are fewer galaxies at high redshifts simply because of the
hierarchical nature of galaxy assembly.

In summary, the HIMF provides a strong constraint on outflow models.
The agreement of the ezw model predictions with the latest observed
HIMF from ALFALFA represents a non-trivial success that has not
previously been attained in hierarchical models of galaxy formation.
The vzw model fares slightly worse but may still be within the
overall uncertainties, while the constant wind and no wind cases
fare poorly against the observed data.  More broadly, this indicates
that including a well-motivated model for galactic outflows enables
hierarchical structure formation models to produce an HIMF that is
in very good agreement with observations down to fairly low \HI\
masses.

\section{HI richness}\label{sec:HIfrac}

\begin{figure}
\vskip -0.5in
\setlength{\epsfxsize}{0.6\textwidth}
\centerline{\epsfbox{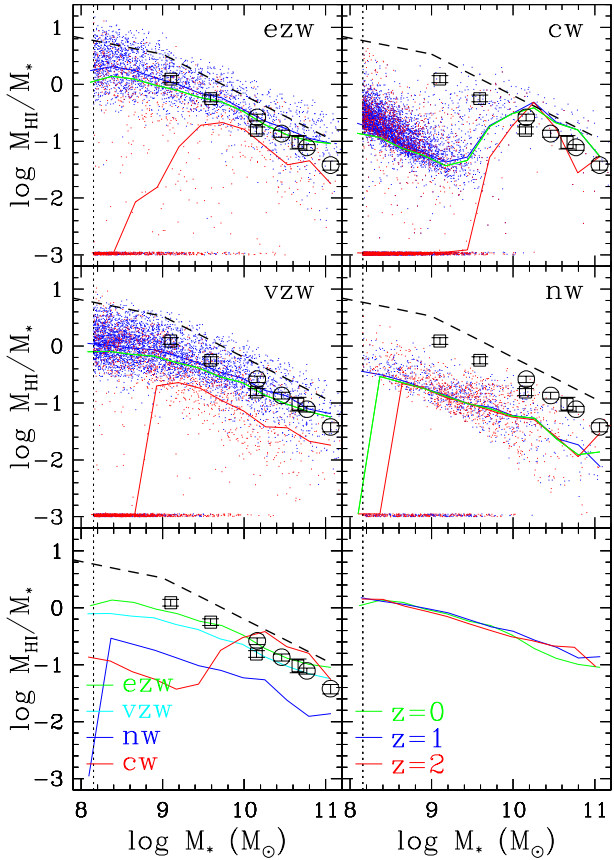}}
\vskip -0.3in
\caption{\HI\ richness ($\equiv M_{\rm HI}/M_*$) in galaxies from our
four wind models:  ezw (upper left), vzw (middle left), 
cw (upper right), and nw (middle right).  Points show individual
galaxies at $z=0$, with central galaxies shown in blue and satellites
shown in red.  A running median for all galaxies is shown in green, while
medians for the centrals and satellites are shown in blue and red, respectively.
Circles show \HI\ fractions observed in the GASS survey of nearby galaxies 
\citep{cat10}, squares show results from the Herschel Redshift Survey~\citep{cor11}, and
the dashed line shows an approximate fit to the locus traced by the 
\HI-selected ALFALFA survey~\citep{hua12}.  The bottom left panel shows
a comparison of all four wind models with observations at $z=0$, while
the bottom right panel shows the evolution of the median \HI\ richness
with redshift out to $z=2$ in the ezw model.
}
\label{fig:HIfrac}
\end{figure} 

The \HI\ richness, i.e. the \HI\ mass relative to the stellar mass,
provides a complementary characterisation of the \HI\ content of
galaxies.  As we argued earlier, both the \HI\ and stellar masses
are reasonably robust predictions of our models,
and hence the \HI\ richness should provide a meaningful
discriminant between models.

Figure~\ref{fig:HIfrac} shows the \HI\ richness in our simulated
galaxies, as a function of stellar mass.  The top four panels show
our four wind models:  ezw (upper left), vzw (middle left), constant
(upper right), and no winds (middle right).  The points in each
panel show individual galaxies at $z=0$, with blue depicting central
galaxies and red depicting satellites.  Galaxies with zero \HI\
content are plotted along the bottom of each panel; they are almost
exclusively satellite galaxies, which we will examine further in
\S\ref{sec:HIenv}.  A binned median of $\log{M_{HI}/M_*}$ for all
galaxies is shown as the green line.  We show errorbars on the
median corresponding to the $1\sigma$ spread for the galaxies within
each mass bin.  We also separately show binned medians for the
central and satellite populations.  The vertical dotted line shows
our galaxy stellar mass resolution limit.

For comparison, mean $M_{\rm HI}/M_*$ observations from the GALEX
Arecibo SDSS Survey (GASS) are shown as the large circles~\citep{cat12}.
GASS has measured \HI\ in a stellar mass-limited sample down to
$M_*\sim 10^{10} M_\odot$.  At lower masses, we show the results
from the Herschel Reference Survey (HRS) by \citet[open squares]{cor11},
which uses literature \HI\ data but is approximately stellar
mass-complete.  We also show the fit to results from the ALFALFA
survey by \citet[dashed line]{hua12}; since this is \HI-mass selected,
it is biased towards more \HI-rich galaxies, as is evident when
compared to the $M_*$-selected data.  Our simulation results are
most straightforwardly comparable to $M_*$-limited samples.

All wind models broadly predict that \HI\ richness is anti-correlated
with $M_*$, as observed.  However, in detail, the models show
distinct differences; for clarity, just the overall medians are
plotted in the lower left panel.  Our ezw model produces a good
agreement with the stellar mass-selected observations down to the
lowest probed masses.  The trend for the vzw model follows ezw, but
shows slightly lowered \HI\ richness particularly in smaller galaxies,
indicating that more mass-loaded galactic outflows in low-mass
galaxies are favoured.  The no-wind case follows the trend of the
data, but is too low by a factor of a few in \HI\ richness, showing
that there is overly efficient conversion of gas into stars in this
model.  Finally, the constant wind model produces a distinct feature
in \HI\ richness, mimicking the feature seen in the HIMF, owing to
wind recycling.  This agrees poorly with the observations, which
have no comparable feature.  Note that more highly mass-loaded
outflows yield {\it higher} \HI\ richness, not lower as one might
naively expect, in part because such outflows suppress $M_*$.

The redshift evolution of \HI\ richness is shown in the lower right
panel of Figure~\ref{fig:HIfrac}, for our ezw simulation.  Despite
the rapid evolution in the HIMF, there is remarkably little evolution
in the \HI\ fraction at a given stellar mass from $z=0\rightarrow
2$.  The slow evolution in \HI\ content contrasts to the much
more rapid evolution observed in the molecular gas
fraction~\citep[e.g.][]{tac10,gea11}, although some of that evolution
may reflect uncertainties in assessing the molecular gas content
from CO emission~\citep[e.g.][]{nar12,bol13}.  Also, the lack of
evolution in \HI\ richness indicates that the evolution in the HIMF
discussed in \S\ref{sec:massfcn} mostly reflects the evolution in
the stellar and/or halo mass functions.  We will see in \S\ref{sec:omegaHI}
that the lack of evolution in global \HI\ content is also a generic
prediction of our models.

\section{\HI\ in Satellite Galaxies}\label{sec:HIenv}

\begin{figure}
\vskip -0.3in
\setlength{\epsfxsize}{0.65\textwidth}
\centerline{\epsfbox{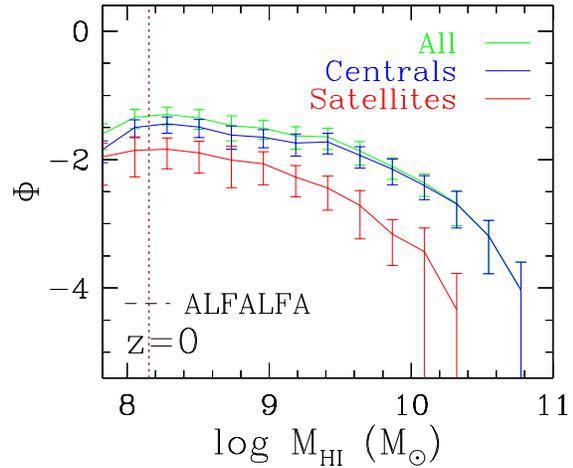}}
\vskip -3.0in
\caption{HIMF from our fiducial ezw simulation (green line, reproduced
from Figure~\ref{fig:massfcn}), separated into centrals (blue) and 
satellites (red).  Centrals dominate the HIMF at all \HI\ masses probed.
}
\label{fig:mf_sat}
\end{figure} 

\begin{figure}
\vskip -0.3in
\setlength{\epsfxsize}{0.6\textwidth}
\centerline{\epsfbox{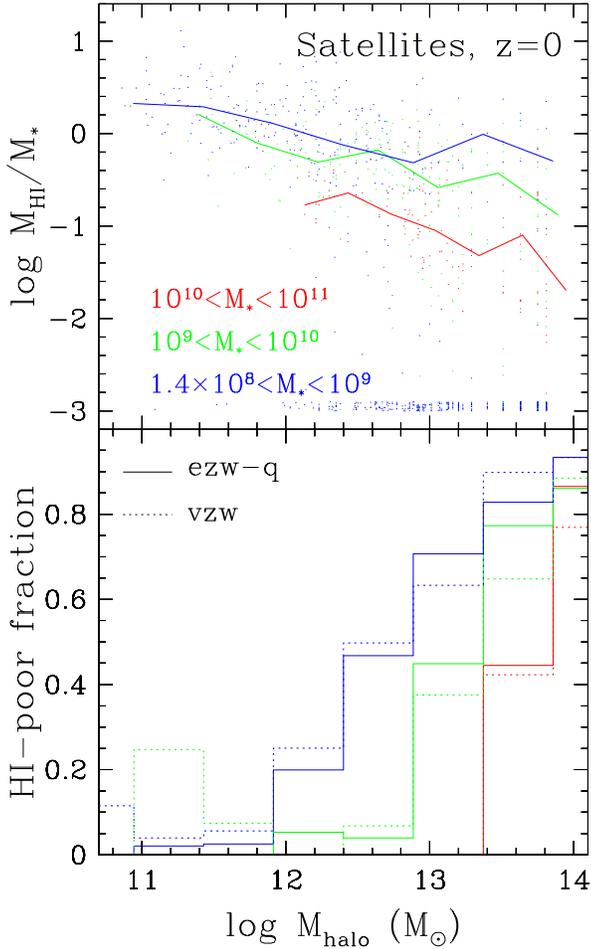}}
\vskip -0.2in
\caption{The top panel shows \HI\ richness ($\equiv M_{\rm HI}/M_*$)
for satellite galaxies vs. halo mass at $z=0$, for our ezw model.  
The satellites are split into low (blue), intermediate (green),
and high (red) stellar mass bins, as indicated.  Satellites with
no \HI\ are plotted at $-3$, with an artificial spread to aid
visualisation.  Lines show a running median of only the galaxies
that have \HI\ (i.e. ignoring the points plotted along the bottom).
The bottom panel plots the fraction of \HI-free
satellites within those same mass bins, colour coded as above.  Solid
lines show results from the ezw model that includes an ad hoc quenching
model at high halo masses, while dotted lines show the equivalent plots
for the vzw model that does not include such quenching.
}
\label{fig:HIsat}
\end{figure} 

The \HI\ content of galaxies is seen to vary substantially with
environment.  In the densest environs such as clusters, ram pressure
can remove \HI\ from infalling galaxies, and it is long known that
cluster galaxies are deficient in \HI~\citep[e.g.][]{hay84}.  Even
in less extreme environments where ram pressure stripping is expected
to play less of a role, the \HI\ content appears to be anti-correlated
with local density~\citep{rob12}; other processes such as strangulation
or harassment may also be playing a significant role.  In general,
such environmental processes are expected to preferentially lower
the \HI\ content of satellite galaxies, particularly those within
larger halos that are expected to have hotter ambient gas.  In this
section we examine how the \HI\ content of satellites varies with
halo mass, which can be considered as a rough proxy for environment.

To begin, in Figure~\ref{fig:mf_sat} we separate the total $z=0$
HIMF in our fiducial ezw simulation (green line) into centrals (blue) and
satellites (red).  Centrals clearly provide the dominant contribution
to the HIMF at all stellar masses probed, being at least three times
more abundant than satellites at any given mass.  Though we do not
show it, this trend holds at all redshifts.  It is also consistent
with the findings from recent semi-analytic models, though at even
lower masses ($M_{\rm HI}\la 10^7 M_\odot$) these models predict
that satellites begin to dominate~\citep[e.g.][]{lag11}.  The basic
trend arises because satellites are less abundant than centrals at
all masses down to $\sim 10^8 M_\odot$~\citep{dav11a}.  As we discuss
next, this is further exacerbated by the trends of \HI\ in centrals
versus satellites.

As is evident from Figure~\ref{fig:HIfrac}, satellites (red points
and curve) have a lower \HI\ richness than central galaxies (blue)
of the same stellar mass.  At the high mass end, the typical
difference in median \HI\ richness between central and satellites
is a factor of $\sim 2-3$, in all the wind models.  Below $M_*\la
10^{9.5} M_\odot$, the satellite \HI\ fraction drops very quickly,
as there are numerous low-mass \HI-free satellites in our simulations
(plotted along the bottoms of the panels).  While the majority of
galaxies of low mass are still centrals~\citep{dav11a}, and hence
the overall median tracks that of central galaxies, it is clear
that low-mass satellites in particular are highly deficient in \HI\
for their mass.  Hence not only are satellites less abundant at a
given $M_*$, they also have less \HI, and hence contribute very
little to the HIMF.

Figure~\ref{fig:HIsat} examines these trends more closely.  The top
panel shows the \HI\ richness of satellite galaxies in our ezw
simulation at $z=0$, divided into three stellar mass bins of
$1.4\times 10^8-10^9 M_\odot$, $10^9-10^{10}M_\odot$, and
$>10^{10}M_\odot$.  Galaxies with \HI\ richness less than $10^{-3}$
(virtually all of which are \HI-free) are plotted along the bottom
at $-3$;  lines show a running median {\it not} including these
\HI-free galaxies.

This plot shows several key trends.  First, at all masses, the
satellites that have \HI\ show a mild trend of being more \HI-rich
in lower mass halos.  Since the halo mass traces stellar mass in
our models~\citep{dav11a}, this basically reflects the fact that
lower stellar mass galaxies have higher \HI\ richness (as seen in
Figure~\ref{fig:HIfrac}).  Similarly, lower-mass satellites are
more \HI\ rich, again following the trend for centrals discussed
in the previous section.  Overall these trends appear to reflect
the \HI\ content of the satellite galaxies when they were still
centrals.

However, a clear difference between central and satellite galaxies
is that there are many more satellites that are devoid of \HI,
particularly at higher halo masses, i.e. the points along the bottom
of the plot appear much more frequently in the more massive halos.
In fact, the distribution of \HI\ richness in satellites appears
to be bimodal.  Even in large halos, there are still some satellites
that have substantial \HI, and their \HI\ content is not grossly
different than that of satellites in smaller halos.  However, the
{\it fraction} of satellites that are \HI\ poor increases sharply
in higher mass halos.

To quantify this, we plot in the bottom panel of Figure~\ref{fig:HIsat}
the fraction of galaxies with \HI\ richness $<10^{-3}$, as a function
of halo mass, divided again into the same three satellite galaxy
stellar mass bins as in the top panel.  Here, we clearly see that
at any halo mass, lower-mass satellites are more likely to have had
their \HI\ content strongly reduced.  Moreover, this trend is a
very strong function of halo mass, with satellites of all masses
being much more likely to be \HI-poor if they lie within a more
massive halo.

The strong bimodality suggests that the process that renders
satellites \HI\ poor happens on a relatively short timescale compared
to the infall timescale into the halo (which is roughly comparable
to the halo's dynamical time of several Gyr).  In future work we
plan to investigate the detailed dynamical processes that remove
the \HI\ from satellites in our simulations, though this will 
likely require some modification to the hydrodynamics algorithm
\citep{age07,rea12,sai13}.

The dotted lines in the bottom panel of Figure~\ref{fig:HIsat} show
the analogous results for our vzw simulation.  In massive halos
where the fraction of \HI-poor satellites is substantial, the outflow
model is identical between vzw and ezw simulations, but the ezw
simulation includes our quenching prescription.  The fact that the
dotted and solid lines are very similar indicate that the trends
seen in the \HI\ content of satellites are not being set by our ad
hoc quenching prescription, but more likely by the fact that halo
masses above $10^{12} M_\odot$ tend to contain much more hot halo
gas that can more strongly impact satellites moving through
it~\citep{ker05,gab11}.  Recall that our quenching prescription
applies only to galaxies with a high velocity dispersion, which is
typically only the central galaxy in the halo, and hence the only
direct impact on smaller satellites would be from the extra heat
being added to the halo gas; evidently this has a minimal effect
on the satellites in our simulations.

What about the central galaxies?  In general, very few of the central
galaxies are devoid of \HI, as seen in Figure~\ref{fig:HIfrac}.
However, although we don't highlight it, there exists a small
population of lower-mass ($M_*\la 10^{10} M_\odot$) centrals that
are \HI-poor.  This is related to what was seen by \citet{gab12},
who found numerous low-mass central galaxies on the red sequence.
These turned out to be galaxies that reside just outside, i.e.
within several virial radii, of more massive halos.  In the spherical
overdensity algorithm we use to identify halos, such galaxies are
identified as centrals, although the influence of the larger galaxy's
halo can extend to well beyond its virial radius~\citep[e.g.][]{mol09}.
Hence these \HI-poor centrals could be impacted by the extended
environment of a nearby larger halo, or else they could be former
satellites whose orbit has taken them outside the nominal virial
radius.

In summary, halo mass plays an increasingly important role in setting
the \HI\ content of satellite galaxies.  This is particularly seen
by the strongly increasing fraction of \HI-poor satellites as a
function of halo mass.  At a given halo mass, low-mass satellites
have a greater chance of having their \HI\ removed.  Satellites
that have not had most of their \HI\ removed lie along similar
relations to satellites in lower-mass halos.  These results suggest
that the process by which \HI\ is removed from satellites in our
simulations acts fairly quickly, and preferentially on smaller
galaxies.  Comparing these predictions to observations can help
constrain such \HI\ removal mechanisms.

\section{\HI\ deficiency}\label{sec:HIdef}

\begin{figure}
\vskip -0.4in
\setlength{\epsfxsize}{0.6\textwidth}
\centerline{\epsfbox{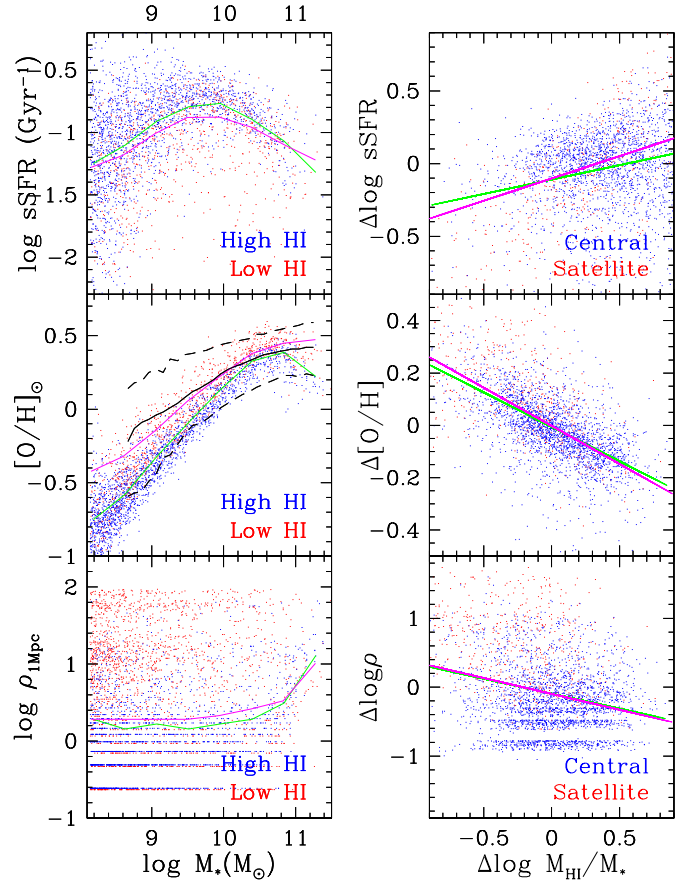}}
\vskip -0.4in
\caption{{\it Left panels:} Specific SFR, metallicity, and environment vs. 
stellar mass (top to bottom) from our ezw simulation at $z=0$, 
with the blue and red points indicating galaxies that have above
and below (respectively) the median \HI\ richness at that given $M_*$.
The green line shows the running median.  The cyan line shows the
median for the vzw model, although the points are not shown.
The black solid line in the middle-left panel plots the median observed
relation from \citet{tre04} with the dashed lines indicating the 95\% contours.
\HI-rich galaxies at a given $M_*$ 
tend to have higher sSFR, lower metallicity, and live in less dense regions.  
{\it Right panels:} The deviation from the median sSFR, metallicity, and
environment (top to bottom) at a given $M_*$ vs. the deviation from the 
median \HI\ richness.  The magenta lines show power law fits to the deviations, 
having slopes of $0.31, -0.26, {\rm and} -0.56$, respectively.
}
\label{fig:HIdeficiency}
\end{figure} 

The \HI\ richness of galaxies is observed to correlate with a variety
of galaxy properties besides stellar mass.  In the previous section
we showed that satellite galaxies in higher mass halos were
increasingly stripped of their \HI.  But even galaxies that still
have substantial \HI\ show correlations of their \HI\ content with
properties such as environment~\citep[e.g.][]{cor11}, and
metallicity~\citep{rob12}.  Here we examine the second-parameter
trends of \HI\ with environment, star formation rate, and metallicity
in our simulations to provide insights into the physical drivers
that establish the \HI\ content of galaxies.

To set a theoretical context for this discussion, we recall the
equilibrium model for galaxy growth as described in \citet{dav12},
which presents a simple physical scenario that can account for the
relationships between key galaxy properties, as well as the scatter
around those relationships.  In the equilibrium model, accretion
onto a galaxy from the cosmic web is fairly quickly processed into
either stars or an outflow, resulting in a slowly-evolving gas
reservoir.  This results in fairly tight relations between stellar
mass and star formation rate~\citep[the so-called main sequence,
e.g.][]{dav08}, gas-phase metallicity~\citep{fin08}, and star-forming
gas content~\citep{dav11b}.  The analytically predicted relations are in good
agreement with full hydrodynamic simulation results.

Stochasticity in the inflow rate causes scatter about these equilibrium
relations.  As described in \citet{dav11b}, an accretion event such
as a minor merger will cause an increase in gas content that raises
the star formation rate, while simultaneously lowering the metallicity.
Conversely, a lull in accretion, or diminished accretion owing to
a galaxy becoming a satellite in a larger halo, will cause the
existing gas to be consumed, resulting in a lower gas content, a
lower SFR, and a higher metallicity.  Hence, as an example, departures
from the mass-metallicity relation are expected to be inversely
correlated with SFR.  Such a trend has been observed~\citep{lar10,man10},
and is known as the fundamental metallicity relation; our simulations
naturally predict this~\citep{dav11b}.  Our models analogously
predict that deviations in the star-forming (i.e. molecular)
gas content will inversely correlate with deviations in
metallicity~\citep{dav11b}.

We now extend this to consider the impact of inflow stochasticity
on the \HI\ reservoir of galaxies.  The left panels of
Figure~\ref{fig:HIdeficiency} show the specific SFR, the metallicity,
and the local galaxy density averaged over 1 Mpc spheres as a
function of stellar mass in our ezw simulation at $z=0$.  Metallicity
here is computed as the SFR-weighted oxygen
metallicity~\citep{fin08,dav11b}.  The observed mass-metallicity
relation (MZR) from SDSS~\citep{tre04} is shown as the solid black
line with contours enclosing 95\% of the simulated galaxies indicated
by the dashed lines.  We have converted these metallicities to solar
units assuming a solar oxygen abundance from \citet{asp09}, namely
$12+\log {\rm [O/H]}_\odot = 8.70$.  The green line shows a running
median in each panel.  We also show as the magenta line the median
for the vzw simulation (without showing the individual points), for
comparison.

The median sSFR drops slowly with $M_*$ at at both small and large
stellar masses, with a peak at $M_*\sim 5\times 10^9 M_\odot$; there
are minimal differences between the ezw and vzw models.  The vzw
line follows that in~\citet{dav11a}, who used lower-resolution
simulations that only probed down to $M_*\sim 10^9 M_\odot$, while
here the turnover is much more apparent.  We note that SDSS
observations do not indicate such a turnover~\citep{sal07}, which
reflects a generic problem in hierarchical models that star formation
in dwarfs peaks too early and is too low today~\citep{wei12}.

The metallicity is tightly correlated and increases with $M_*$ until
$M_*\sim 5\times 10^{10} M_\odot$.  The ezw model produces a steeper
mass-metallicity relation (MZR) than the vzw model, for reasons
discussed in \citet{fin08} and \citet{dav12}.  The metallicity
$Z\propto 1/(1+\eta)$ (in the absence of wind recycling), with
$\eta\propto 1/\sigma\propto M_*^{-1/3}$ for all galaxies in the
vzw model and $\eta\propto 1/\sigma^2\propto M_*^{-2/3}$ for the
ezw model at small masses.  For $\eta\gg 1$, as at small masses,
this toy model yields $Z\propto M_*^{1/3}$ for the vzw model and
$Z\propto M_*^{2/3}$ for the ezw model.  This represents an asymptotic
slope, and since the ezw model follows the vzw model scalings at
high masses, the actual MZR in the ezw model is not as steep as
this, but it is clearly steeper than in the vzw model.  The vzw
model appears to provide a better fit to the SDSS data, although
metallicity measures are sufficiently uncertain that the ezw
prediction cannot be ruled out~\citep{ell08}.  Also, a recent
determination of the MZR from direct metallicity measures by
\citet{and13} suggests a steeper MZR slope at small masses, closer
to $Z\propto M_*^{1/2}$.

The simulated galaxies in Figure~\ref{fig:HIdeficiency} are
colour-coded by \HI\ richness:  the blue points show galaxies that
have a higher-than-median \HI\ richness at that stellar mass, and
the red points show the converse.  It is clear that \HI\ richness
has a strong second parameter correlation with all these properties:
\HI-rich galaxies at a given mass also tend to have a higher sSFR,
a lower metallicity, and a lower local galaxy density.

To reiterate the physical scenario: when accretion happens, it
results in galaxies with higher \HI\ richness than typical.  This
high \HI\ richness also corresponds to a lower-than-normal gas-phase
metallicity.  This suggests that the \HI\ content of galaxies can
be an indicator of recent accretion, as argued from observations
by \citet{mor12}, who show that the low metallicities in \HI-rich
systems most strongly appear at the outskirts of disks.

We can quantify these trends using deviation plots, which we show
in the corresponding right panels of Figure~\ref{fig:HIdeficiency}.
This shows the difference between a galaxy's sSFR, metallicity, and
density and the median value of these quantities at a given $M_*$,
i.e. approximately the green line in the left panels (although in
detail we employ a spline fit to produce more smoothly varying
deviations), versus the deviation in \HI\ richness with its median
value at that $M_*$.  Points in these panels are colour-coded by
centrals (blue) and satellites (red).  The green lines show power-law
fits to these deviations for the ezw model and the magenta lines
show the same for the vzw model (though the individual points are
not shown).

These right panels quantify the second-parameter trends from the
left panels: galaxies with high \HI\ content also have a high sSFR,
a low metallicity, and are in low density environments.  A power
law provides a reasonable fit to the sSFR and metallicity data,
with slopes of 0.31 and $-0.26$, respectively.  The vzw model yields
virtually identical slopes, even though the trends (such as the
MZR) are noticeably different.  This occurs because the trend in
the scatter arises from inflow stochasticity, which has little to
do with the outflows.  The central and satellite galaxies do not
show significantly different trends in sSFR or metallicity.

The metallicity dependence can be compared to recent work by
\citet{rob12}, who looked at the \HI\ deficiency parameter, $DEF$,
relative to the deviation from the expected oxygen abundance for
cluster and field galaxies, finding slopes of $-0.25\pm0.12$ and
$-0.41\pm0.14$, respectively.  Our simulated galaxies are more
comparable to field galaxies, which show a higher slope but are
still within their uncertainties.  \citet{rob12} compared to results
from the simulations in \citet{dav11b}, which yielded a similar
slope.  Although the definition of DEF differs from the deviation
we plot here, the qualitative agreement in the trend indicates that
the basic physical model of a slowly-evolving equilibrium in gas
content and metallicity appears to be broadly consistent with the
observations.

Finally, we consider the environment in the bottom two panels of
Figure~\ref{fig:HIdeficiency}.  The median local galaxy density is
independent of mass until the very highest mass systems in our
volume at $M_*\ga 5\times 10^{10} M_\odot$.  Unlike in the sSFR and
metallicity deviation plots, where the centrals and satellites do
not significantly deviate from one other, the environment deviation
plot (bottom right) markedly separates the central and satellite
galaxies.  This environmental dependence arises because the environment
can impact the inflow rate into satellites.  High density regions
associated with halos with masses $\ga 10^{12}M_\odot$ will contain
substantial amounts of hot gas~\citep{ker05,ker09,gab12} that can
retard accretion~\citep{dek06,dav12}.  Hence galaxies in such
regions, particularly satellites, will have lower accretion rates
compared to field galaxies at the same mass, and hence less \HI.
Overall, the full galaxy sample trends towards having less gas-rich
galaxies in denser environments, but this is driven almost entirely
by the satellites.  Fitting a power-law formally yields a slope of
$-0.56$ (shown as the magenta line), though a power law does not
appear to be a particularly good fit.

In summary, the correlated deviations between sSFR, metallicity,
and \HI\ content provide quantitative constraints in the way in which
galaxies oscillate about their equilibrium relations owing to
stochastic accretion.  The resulting deviation slope reflects the
correlation between the instantaneous gas inflow rate, its conversion
to neutral hydrogen, and the infalling metallicity, as mediated by
the local environment.  Such constraints provide valuable discriminants
between different physical scenarios for galaxy growth.  The
consistency with available observations, albeit preliminary and
qualitative, indicates that the stochasticity in the inflow as
expected from cosmological accretion is able to broadly explain the
second parameter trends observed in the relationship between \HI\
content and other physical galaxy properties.

\section{$\Omega_{\rm HI}$ evolution}\label{sec:omegaHI}

\begin{figure}
\vskip -0.5in
\setlength{\epsfxsize}{0.65\textwidth}
\centerline{\epsfbox{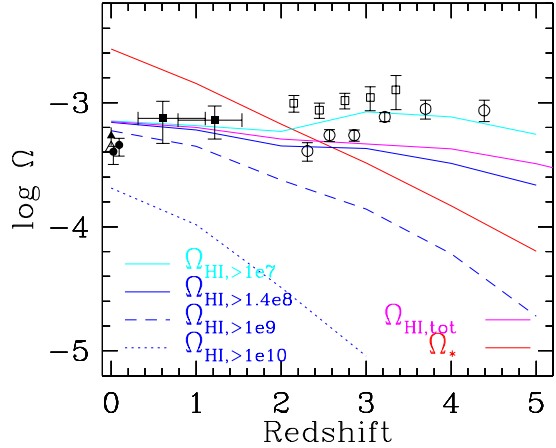}}
\vskip -3.0in
\caption{$\Omega_{\rm HI}$, the fraction of cosmic mass density in
\HI, from $z=5\rightarrow 0$ from our ezw simulation.  
The blue solid line shows $\Omega_{\rm HI}$ from all resolved galaxies
($M_*>1.4\times 10^8 M_\odot$)
the blue dashed and dotted lines show $\Omega_{\rm HI}$ in galaxies with 
$M_*>10^9 M_\odot$ and 
$M_*>10^{10} M_\odot$, respectively.  The cyan line shows $\Omega_{\rm HI}$ from
extrapolating a Schechter function fit to the HIMF at each redshift
down to $M_{HI}=10^7 M_\odot$.  The magenta line shows the total \HI\ mass
density in the entire volume.  For comparison, the red line shows the mass 
density in stars in this model.  Observations are shown 
at $2\la z\la 4.5$ from \citet{pro09} (open circles),
at $2\la z\la 3.5$ from \citet{not12} (open squares),
at $0.5\la z\la 2$ from \citet{rao06} (filled squares),
at $0.02\la z\la 0.1$ from \citet{del13} (filled circles)
and at $z=0$ from \citet{hay11} (open triangle) and 
\citet{bra12} (filled triangle).
}
\label{fig:omegaHI}
\end{figure} 

Most neutral hydrogen in the cosmos exists in and around galaxies,
between the regime where self-shielding happens in the outskirts
of disks, to where gas becomes molecular-dominated within star-forming
regions.  While it is currently infeasible to probe the evolution
of \HI\ in galaxies via 21cm emission to high redshifts\footnote{This
is a key goal of the upcoming LADUMA survey using the MeerKAT
array~\citep{hol11}}, other avenues have been employed to measure
cosmic \HI\ evolution, such as the abundance of damped \lya\ systems
(DLAs). In this section we examine the evolution of the cosmic \HI\
density in our simulations directly from the galaxy population, and
compare it with various observational determinations.

Figure~\ref{fig:omegaHI} shows the evolution of $\OHI$ from
$z=5\rightarrow 0$ in our ezw model, for all galaxies above our
approximate \HI\ mass resolution limit of $M_{\rm HI}>1.4\times
10^8 M_\odot$ (solid blue line), as well as for galaxies with $M_{\rm
HI}>10^9 M_\odot$ (dashed) and $M_{\rm HI}>10^{10} M_\odot$ (dotted).
For comparison, we also show the evolution of the stellar mass
density $\Omega_*$ as the red line.  A sample of observations are
also shown, as detailed in the caption.  Observations of $\OHI$ at
$z>2$ from \citet{pro09} and \citet{not12} are computed from DLA
abundances, while at $0.5<z<2$ the data is from \citet{rao06} who
used strong \ion{Mg}{ii} systems as a proxy for DLAs.  Low-redshift
($z\la 0.1$) data comes from 21cm emission surveys, including
\citet[ALFALFA]{hay11}, \citet{bra12}, and \citet{del13}.  Note
that our prescription for determining self-shielding calibrated to
radiative line transfer simulations as described in \S\ref{sec:sims}
does reasonably well predicting $\OHI$ at $z=0$; we do not adjust
any parameters to match this data as was done in \citet{pop09}.

Our simulations predict that $\OHI$ is remarkably constant from
$z=5\rightarrow 0$, essentially unchanging to within 50\% over these
12 billion years.  This is in contrast to the dramatic increase of
$\Omega_*$ over that same time interval; over 98\% of all stars in
this model form since $z=4$, with more than 80\% since $z=2$~\citep[in
broad agreement with recent observational estimates; e.g.][]{lei12}.
It is also quite different than the rapid evolution in cosmic star
formation rate density~\citep[e.g.][]{hop06,far07} and molecular
gas fraction~\citep{tac10,gea11,tac12}, which both increase by an
order of magnitude or so out to $z=2$~\citep[though see][who argue
for less growth in the observations owing to variations in the
CO-to-H$_2$ conversion factor]{nar12}.  This emphasises that \HI\
represents a transient reservoir in the journey of gas from the
ionised IGM to stars forming deep within galaxies, and is not
directly proportional to the amount of stars being formed or to the
amount of molecular gas present.  This cautions against over-interpreting
quantities such as the ``\HI\ star formation efficiency", i.e. the
SFR divided by the \HI\ mass, since the two quantities are not
directly related.

Subdividing $\OHI$ into different \HI\ mass bins, we see that at
$z=0$, most of the cosmic \HI\ is in galaxies with $10^9<M_{\rm
HI}<10^{10}M_\odot$.  Going back in time, \HI\ shifts towards lower
mass galaxies; by $z=2$, only half the \HI\ is in $M_{\rm
HI}>10^{9}M_\odot$ systems.  In contrast, only a very small portion
of the cosmic \HI\ ever resides in galaxies with $M_{\rm HI}>10^{10}
M_\odot$, since higher mass galaxies quickly drop off in their \HI\
richness (Figure~\ref{fig:HIfrac}).  Thus quenching mechanisms,
which primarily affect high-mass galaxies~\citep[e.g.][]{gab12},
are expected to have almost no effect on $\OHI$.  We also show, as
the magenta line, the total \HI\ mass density in the entire simulation
volume.  The difference between the solid magenta and blue lines
reflects \HI\ that is not within resolved galaxies (including the
diffuse IGM).  This extra contribution is a couple of a percent at
low redshift but rises to $\sim 50\%$ at $z=5$, which again reflects
the increasing contribution of small (unresolved) galaxies to the
global \HI\ budget at high redshifts.

Observations of $\OHI$ likewise indicate very little evolution from
$z\sim 4\rightarrow 0$.  Different observational tracers generally
agree to within a factor of two, and together indicate
essentially no change in $\OHI$ for the past 12 billion years, in
broad agreement with our predictions.  Our simulation begins to
underpredict $\OHI$ at $z\ga 3$ when compared with the DLA data.
The discrepancy could be physical or numerical.  A physical explanation
could be that lower metallicity galaxies at early epochs will tend
to have less efficient conversion of their atomic gas to molecular
gas owing to lower cooling rates and harder interstellar radiation,
and hence our locally-calibrated prescription for $R_{\rm mol}$
from \citet{ler08} may not be appropriate (see \S\ref{sec:res}).

A numerical explanation for the discrepancy may be that we do not
fully resolve galaxies with $M_*<1.4\times 10^8M_\odot$, and there
may be substantial contributions to $\Omega_{\rm HI}$ from the very
smallest galaxies.  At low-$z$, the low mass end of the HIMF is
fairly shallow, so the expected contribution from lower mass galaxies
is small.  But at higher redshifts, Figure~\ref{fig:massfcn} shows
that the slope becomes substantially steeper, meaning that the
additional contribution from unresolved systems could be large.

We can crudely correct for this by fitting a Schechter function to
the HIMF at each redshift as we did in \S\ref{sec:massfcn}, and
then integrating down to some chosen lower mass limit.  As an
illustration, we integrate down to $M_{\rm HI}=10^7 M_\odot$, which
is around the lowest mass observable at low-$z$ in large surveys
such as ALFALFA.  The result of this Schechter fit extrapolation
to $M_{\rm HI}=10^7 M_\odot$ is shown as the cyan line in
Figure~\ref{fig:omegaHI}.  This results in a negligible correction
at low-$z$, but at higher redshifts the correction can be up to a
factor of three, which agrees better with the high-$z$ DLA results.
We caution that this exercise is intended to be illustrative, since
our $M_{\rm HI}$ limit was chosen rather arbitrarily, and it may
not correspond to the effective \HI\ masses probed by DLA systems.
Furthermore, our Schechter function fits become increasingly uncertain
at higher redshifts, owing to the lack of dynamic range in our
simulations.  Nonetheless, this illustrates that plausible corrections
down to lower \HI\ masses can bring our predicted $\Omega_{\rm HI}$
into better agreement with the observations by preferentially
increasing the high-redshift \HI\ mass density.  Given the crudeness
of our prescriptions for determining \HI\ content, agreement at
this level is encouraging.

In summary, our simulations generally predict a very slowly-evolving
cosmic \HI\ mass density, in broad agreement with the observations.
However, we note that the HIMF actually evolves rather considerably
(Figure~\ref{fig:massfcn}), with many more low-$M_{\rm HI}$ galaxies
and fewer high-$M_{\rm HI}$ ones at high $z$.  It is something of
a coincidence that these two variations roughly cancel when summed
to give the global \HI\ mass density.  Nonetheless, this emphasises
the transient nature of the \HI\ reservoir around galaxies, which
does not build up hierarchically in the way that the stellar mass
does, but instead responds to evolution in the ionising background
and the ISM physical conditions.

\section{Numerical Robustness}\label{sec:res}

\subsection{Resolution Convergence}

\begin{figure}
\vskip -0.5in
\setlength{\epsfxsize}{0.65\textwidth}
\centerline{\epsfbox{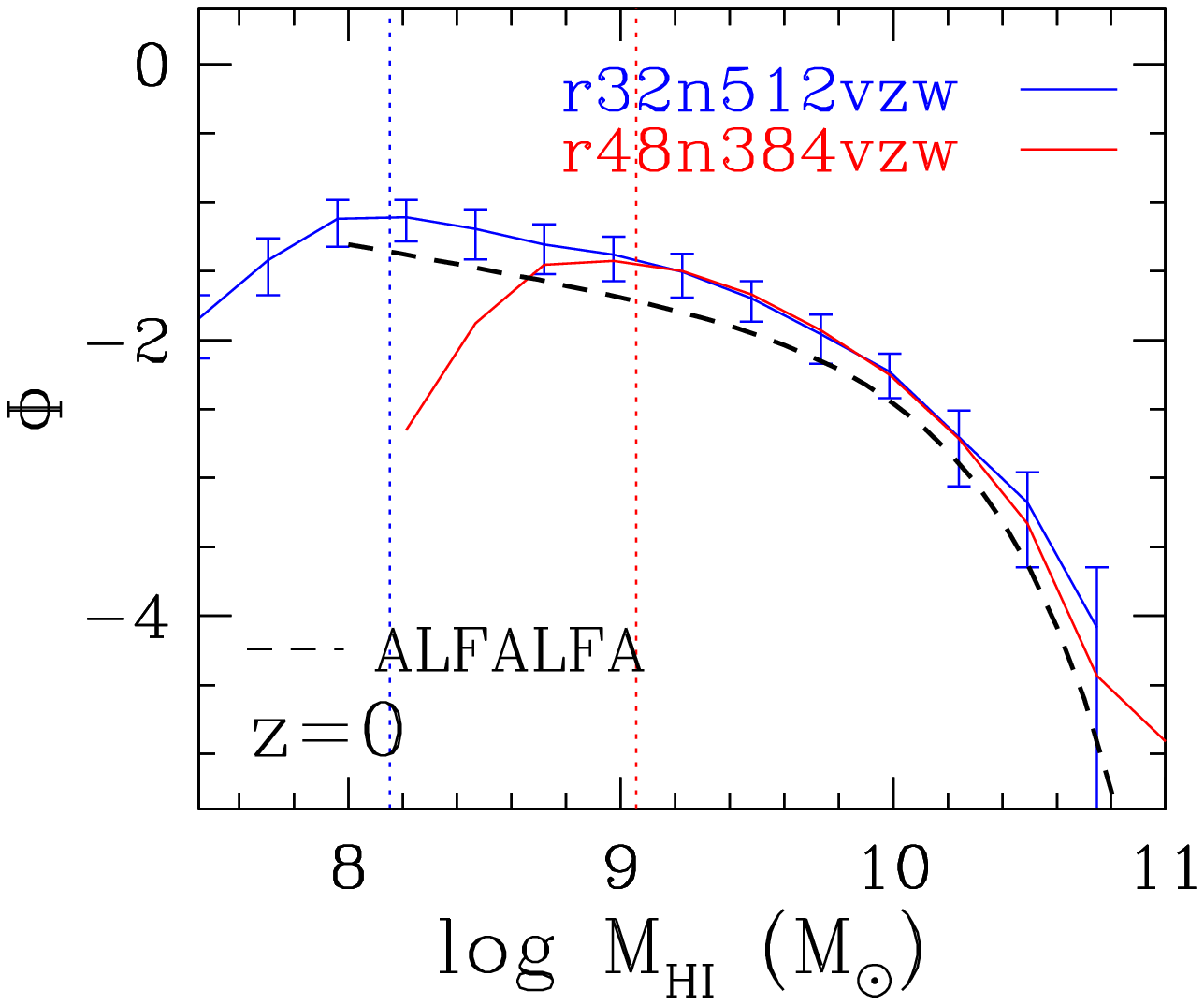}}
\vskip -3.5in
\setlength{\epsfxsize}{0.65\textwidth}
\centerline{\epsfbox{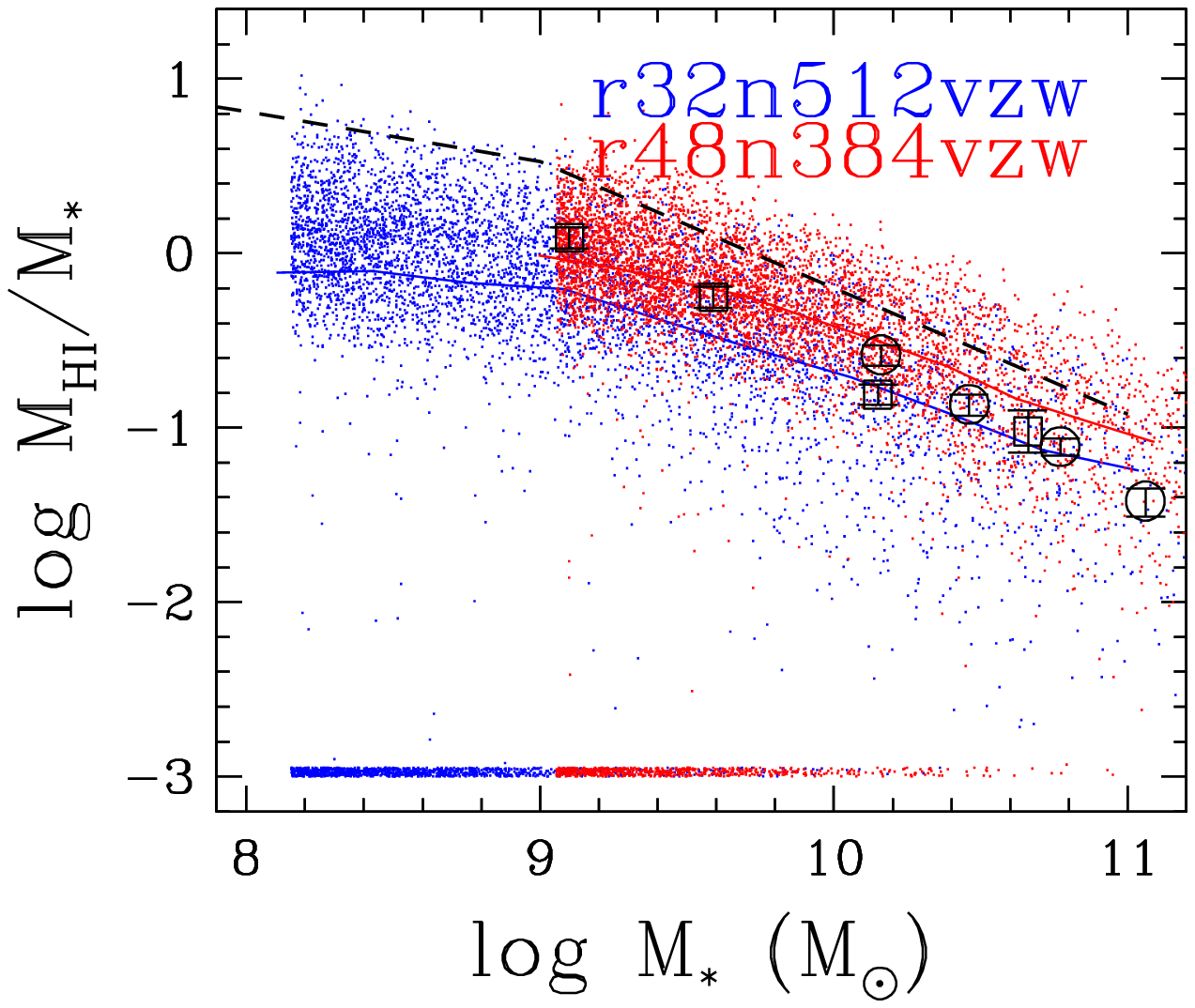}}
\vskip -3.0in
\caption{Comparison of the \HI\ mass function (top panel)
and \HI\ richness (bottom panel) in our fiducial $32\hmpc$,
$2\times 512^3$ particle simulation (green) compared to a
$48\hmpc$, $2\times 384^3$
particle simulation (red) with $8\times$ poorer mass resolution.
Resolution convergence is good for the HIMF, while the \HI\ richness
is decreased in the higher-resolution simulation, but is still within the
formal uncertainties. 
}
\label{fig:HIres}
\end{figure} 

In this study we have employed some of the highest-resolution
hydrodynamic simulations of random cosmological volumes ever evolved
down to $z=0$.  Nonetheless, it is important to assess whether our
resolution is sufficient to robustly predict the \HI\ properties
of galaxies.  Here we examine our two most basic \HI\ statistics,
namely the \HI\ mass function and the \HI\ richness vs. stellar
mass, in a simulation that has $2\times 384^3$ particles in a
$48\hmpc$ volume (r48n384), described in our previous
work~\citep[e.g.][]{opp10,dav11a,dav11b}.  This simulation has 8
times poorer mass resolution and $2$ times poorer spatial resolution,
albeit in a volume that is $3.4$ times larger.  We use the identical
momentum-driven wind (vzw) and cooling model in both simulations\footnote{
We use the vzw wind model because we already have the r48n384vzw
simulation in hand, and there are usually only minor differences
between vzw and our now-favoured ezw model.}.  We also renormalise
the metagalactic background by a flux factor of 1.5, as computed
in \citet{dav10}.

Figure~\ref{fig:HIres} shows the \HI\ mass function (top) and the
$M_{HI}/M_*$ ratio vs. $M_*$ (bottom) in these two simulations,
r32n512vzw in green and r48n384vzw in red.  These are analogous to
Figures~\ref{fig:massfcn} and \ref{fig:HIfrac}, respectively, and
the observations are plotted as in those figures.

The HIMF (top panel) shows very good resolution convergence, at
least over this somewhat modest range in mass resolution probed
here.  The two mass functions are essentially identical until below
the resolution limit (indicated by the colour-coded dotted line),
where the low-resolution simulation starts to become significantly
incomplete.  This also explicitly demonstrates that our chosen \HI\ mass
resolution limit is appropriate when considering the HIMF.

For the \HI\ fraction, the resolution convergence is not quite as
good.  The running median lines for the high resolution simulation
lies slightly below the lower-resolution one, indicating that the
\HI\ richness decreases with increasing resolution.  The effect is
roughly $\sim 50\%$ over this factor of two in spatial (or factor
of eight in mass) resolution, and suggests that even higher resolution
simulations are required to fully assess the resolution convergence.

\begin{figure}
\vskip -0.3in
\setlength{\epsfxsize}{0.55\textwidth}
\centerline{\epsfbox{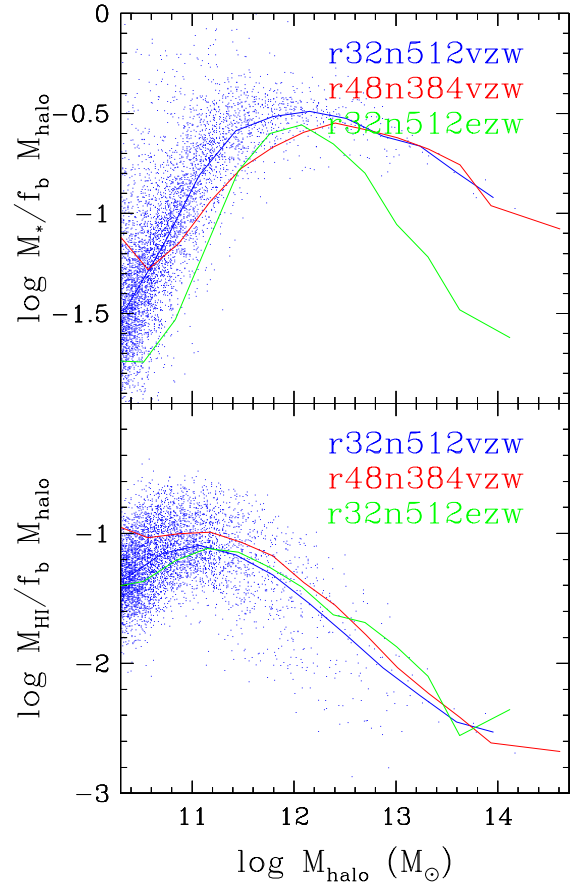}}
\vskip -0.2in
\caption{The mass fraction of stars (top panel) and \HI\ (bottom) 
relative to the expected baryonic mass in the halo ($=f_b M_{\rm halo}$,
where $f_b=0.164$ is the baryon fraction), vs. $M_{\rm halo}$,
for central galaxies in our high-resolution vzw simulation (r32n512vzw, blue), 
low-resolution vzw simulation (r48n384vzw, red),
and our fiducial ezw simulation (r32n512ezw, green).
Increasing the resolution yields higher stellar masses and lower
\HI\ content.
}
\label{fig:HIhalo}
\end{figure} 

To examine the origin of the lack of convergence, we show in
Figure~\ref{fig:HIhalo} the stellar (top) and \HI\ (bottom) mass
relative to the halo baryonic mass (i.e. the halo mass multiplied
by the baryon fraction $f_b$), as a function of halo mass.  We
consider only central galaxies since satellite masses are less
correlated with their host halo masses.  We plot the higher-resolution
vzw simulation in blue, with points, while for the lower-resolution simulation
we plot only the running median in red.  For reference we also plot
our fiducial ezw simulation in green, showing that it lowers the stellar
baryon fraction at both high and low halo masses but that it has a modest
impact on the \HI\ halo fraction at a given halo mass, relative to
vzw.

At a given halo mass, the stellar mass is increased and the \HI\
mass is decreased at higher resolution.  Both effects contribute
roughly equally to lowering the \HI\ richness.  It appears that
higher-resolution simulations are more effective at converting their
\HI\ into stars, possibly because the increased resolution enables
faster cooling through the \HI\ regime.  This suggests that even
higher resolution simulations are needed to properly converge both
these quantities, though the qualitative trends and interpretations
are not impacted by resolution.

Overall, the resolution convergence for \HI\ properties is good but
not ideal.  For a given $M_*$ the \HI\ content is very robust, but
the number density of galaxies at a given \HI\ mass is less well
converged.  For that reason, $\Omega_{HI}$ is also slightly uncertain,
as there is a $\sim 30\%$ difference between the $z=0$ value predicted
by the high and low resolution simulations.  We do not yet know
whether these differences reflect a resolution effect or a volume
effect; we would need a wider suite of simulations to assess these
issues.  While these discrepancies are not trivial, they do not
significantly affect the main conclusions of this paper.

\subsection{Molecular Gas Prescription}

Our prescription for computing the molecular gas fraction relies
on an empirical calibration from The \HI\ Nearby Galaxy
Survey~\citep[THINGS;][]{ler08} fit to their combined spiral sample.
It does not explicitly include any dependence on metallicity or
redshift.  However, there are good theoretical reasons to believe
that the conversion of atomic to molecular gas should depend on
metallicity, since dust is an important catalyst for $H_2$ formation,
and possibly on redshift since the typical interstellar medium
properties in high redshift galaxies are different than those today.
Indeed, several groups have developed theoretical models to investigate
this transition, and particularly its dependence on
metallicity~\citep{kru08,gne11}.

\begin{figure}
\vskip -0.2in
\setlength{\epsfxsize}{0.45\textwidth}
\centerline{\epsfbox{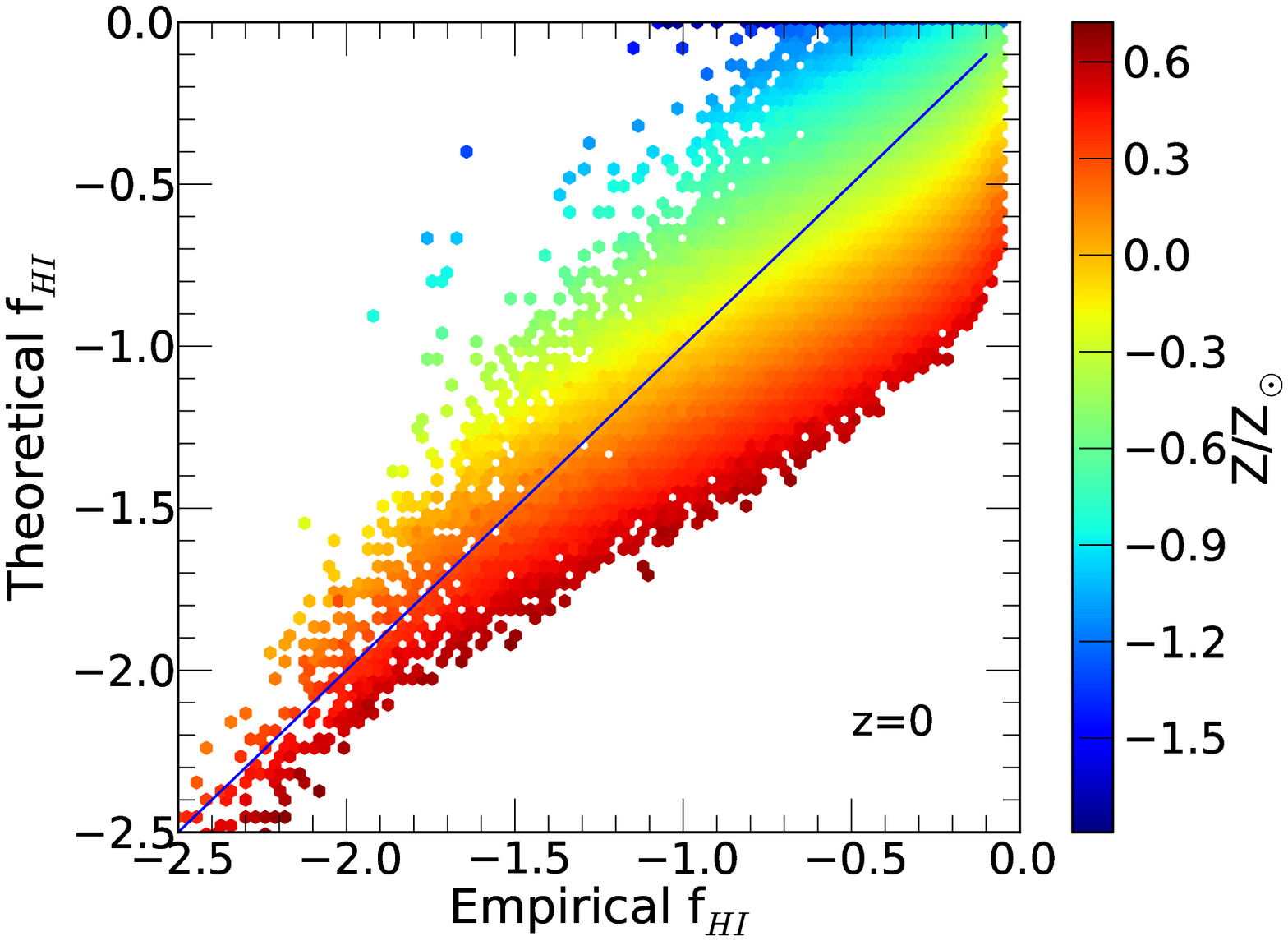}}
\setlength{\epsfxsize}{0.45\textwidth}
\centerline{\epsfbox{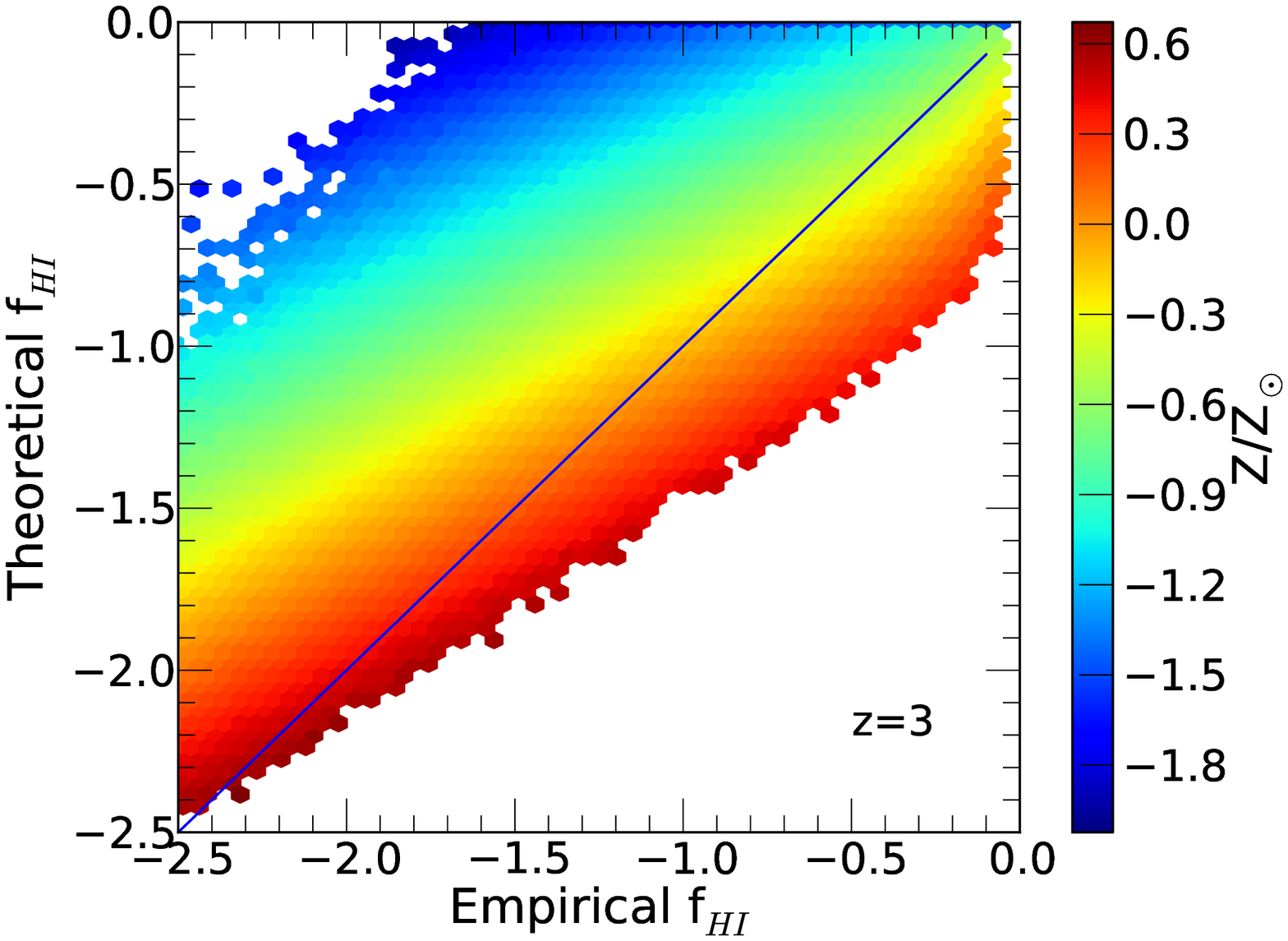}}
\caption{ Comparison of star-forming gas particle neutral fractions 
computed using the observationally-derived empirical pressure relation 
for the $H_2$ content from \citet{ler08}
vs. the theoretically-derived relation from \citet{kru11}, at
$z=0$ (top panel) and $z=3$ (bottom).  Particles are binned and colour-coded
by the mean metallicity within each bin.  The theoretically-determined
neutral fraction is higher for lower metallicity gas, and there
is more such gas at higher redshifts.
}
\label{fig:H2model}
\end{figure} 

\begin{figure}
\vskip -0.5in
\setlength{\epsfxsize}{0.65\textwidth}
\centerline{\epsfbox{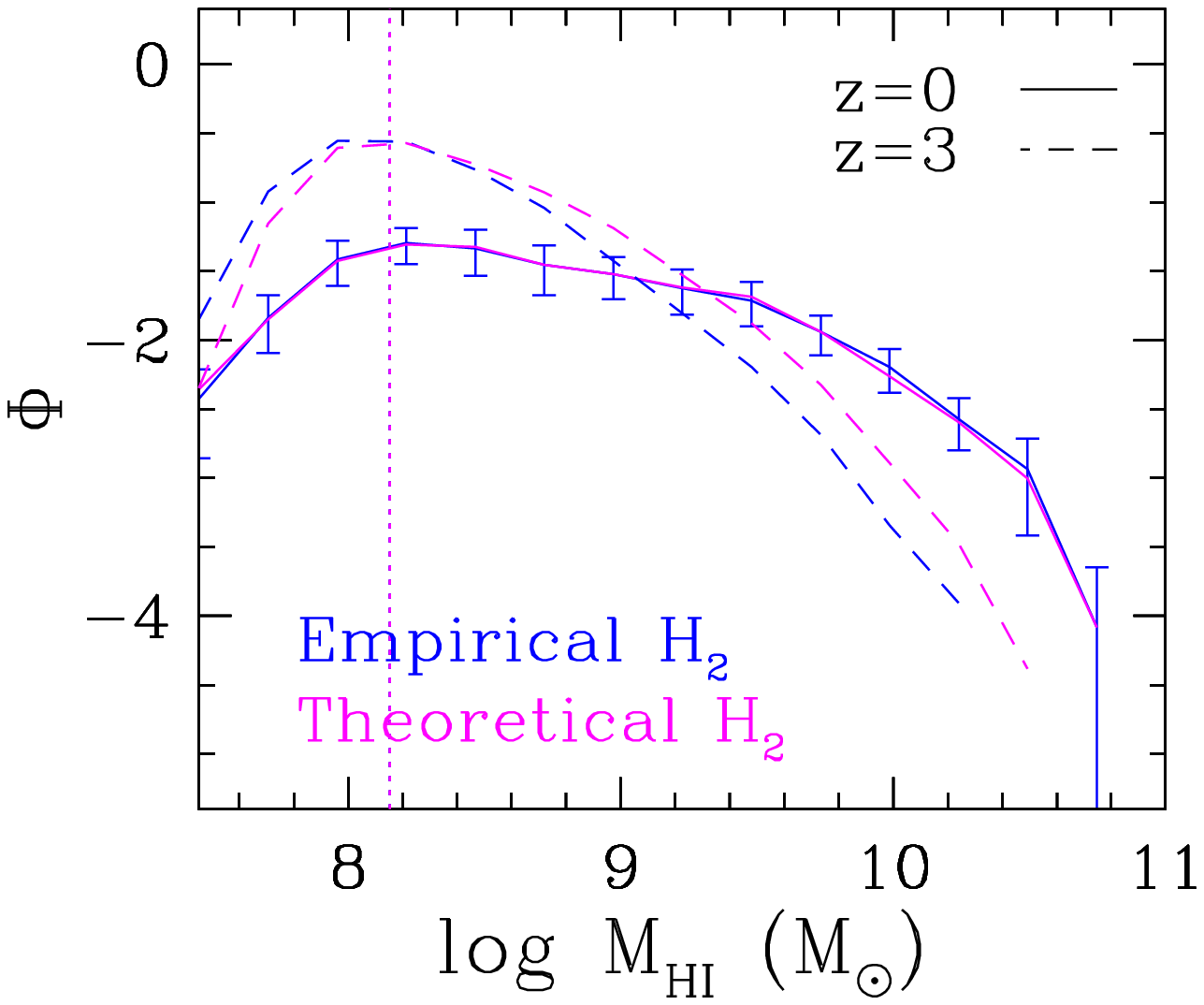}}
\vskip -3.0in
\caption{Comparison of the \HI\ mass function using the observationally-
derived empirical pressure relation for $H_2$ fraction from \citet{ler08}
(blue lines),
vs. the theoretically-derived relation from \citet{kru11} (red lines).
Solid lines show results at $z=0$, where there is essentially no
difference between these prescriptions.  Dashed lines show $z=3$
results, where the metallicity dependence in the theoretical relation
generally yields a higher neutral fraction.
}
\label{fig:mfH2}
\end{figure} 

In this section we employ the analytic prescription developed by \citet{kru08}
to separate atomic and molecular hydrogen, and assess its impact on the 
HIMF.  The prescription follows that detailed in \citet{kru11}:
\begin{equation}
f_{H2} = 1 - \frac{0.75 s}{1+0.25 s }
\end{equation}
where 
\begin{equation}
s = \frac{ln(1+0.6\chi+0.01\chi^2)}{0.6\tau_c},
\end{equation}
\begin{equation}
\chi = 0.756 (1 + 3.1 Z^{0.365}),
\end{equation}
and
\begin{equation}
\tau_c = \Sigma \sigma_d/\mu_H.
\end{equation}
Here, $Z$ is the metallicity in solar units, $\sigma_d$ is the cross-section
for dust, which we assume to be equal to $Z \times 10^{-21}$ cm$^2$, 
$\mu_H=2.3\times 10^{-24}$~g is the mean mass per H nucleus, and $\Sigma$ is 
the gas
surface density that we take to be the gas density times the SPH smoothing
kernel size.  This prescription is applied to each star-forming gas particle
since molecular gas is only expected to form substantially in star-forming
regions; the neutral fraction in non-star forming gas is not altered and is
computed as described earlier (\S\ref{sec:sims}).

Figure~\ref{fig:H2model} shows a scatter plot, colour-coded by
metallicity, of the neutral fraction computed using our empirical
method from \citet{ler08} versus the theoretical method from
\citet{kru11}, at $z=0$ (top) and $z=3$ (bottom).  As expected,
lower metallicity gas has a higher neutral fraction in the theoretical
prescription, showing an order of magnitude change in the neutral
fraction at a given ISM pressure over the two-dex span in metallicity.
At $z=0$, there is little overall systematic shift in the neutral
fraction, indicating that the theoretical method is well-calibrated
to the empirical one for local solar-metallicity spirals from
\citet{ler08}.  However, at $z=3$ where there is more low-metallicity
star-forming gas, the theoretical method yields systematically
higher neutral fractions.

Figure~\ref{fig:mfH2} shows the HIMF using the empirical (blue) and
theoretical (magenta) methods to compute molecular fractions.  The
solid lines show the results at $z=0$.  Although there is a systematic
metallicity dependence as shown in the previous plot, the overall
resulting HIMF at $z=0$ shows essentially no difference between the
two methods, even as a function of mass.  Note that the range in
metallicities from our most massive to least massive galaxies spans
over a dex (see Figure~\ref{fig:HIdeficiency}).  Nonetheless, the
region from where most of the neutral gas originates is typically
well outside the region of vigorous star formation that determines a galaxy's
metallicity.  In these regions, typically at densities of $\la
1$~cm$^{-3}$, the impact of metallicity on the molecular gas content
is evidently not significant.

In contrast, at $z=3$ the HIMF shows a systematic offset towards
higher \HI\ content using the theoretical method.  This arises
because our simulations show a mild evolution in galaxy metallicity
of roughly a factor of two at a given stellar mass out to these
redshifts~\citep{dav11b}, which is consistent with observations out
to $z\sim 2.5$~\citep[e.g.][]{erb06}, although higher redshift data
may suggest a more rapid dropoff~\citep[though see also
\citealt{ric11}]{man10}.  Still, the HIMF is only mildly higher
using the metallicity-dependent theoretical prescription, with a
maximum increase in \HI\ mass of $\sim\times 1.5$.  We conclude
that there is a modest uncertainty in our predictions owing to which
$H_2$ prescription employed, and it primarily affects higher-$z$
predictions.  For instance, it could help to reconcile the predicted 
$\Omega_{\rm HI}(z)$ with the DLA-derived values at $z\sim 3$.

\subsection{Hydrodynamics Methodology}

Our \gad\ simulations employ the entropy-conserving (EC) formulation
of SPH~\citep{spr05}.  While this formulation has the advantage of
simultaneously conserving entropy and energy, it has been shown to
have significant deficiencies particularly when surface instabilities
are present~\citep[e.g.][]{age07}, in that it appears to add an
artificial surface tension that prevents the disruption of cold
clumps moving through a dense medium.  This could influence \HI\
content by, for instance, suppressing gas stripping in infalling
satellites.  Various newer formulations of SPH have been developed
that improve this aspect~\citep[e.g.][]{rea12,sai13,hop13}, but
these are just now being incorporated into full cosmological codes.
Furthermore, all forms of SPH tend to be overly dissipative in shear
flows, despite explicit attempts to suppress shear viscosity~\citep{bal89}.
Hence galaxy disks in SPH tend to be overly condensed~\citep{tor12}
barring the inclusion of strong feedback~\citep{bro12}, which could
affect the conversion of \HI\ to H$_2$ within the disk.  These
issues beg the question whether \HI\ results using EC-SPH might be
severely compromised.

Regarding disk sizes, it is worth noting that our simulations
do possess the sort of strong feedback that results in larger disks
that are more consistent with observations, as we showed using
high-resolution cosmological zoom simulations at $z\sim 2$~\citep{ang13}.
Nonetheless, disk radii in the cosmological simulations we use here are
still somewhat smaller than observed at $z=0$, though not nearly
as far off as in simulations without outflows.  It is unclear
whether this has a substantial effect on the atomic to molecular
transition, because our prescription relies on the ISM pressure
that is determined locally from the subgrid two-phase ISM
model~\citep{spr03b} rather than from the global properties of the
disk.  Even when we use the theoretical prescription described in
the previous section, the surface density is calculated from
individual particles' gas densities and smoothing lengths, rather
than global disk properties.  Nonetheless there may be some cumulative
effect from the improper density profile of the disk.  It is difficult
to assess this directly without either doing much higher resolution
zoom simulations, in which case it is more difficult to compile
statistics, or else using a different code such as Arepo~\citep{tor12},
in which case one gets disks with larger radii but in other
ways still quite unlike observed disks.

Regarding the issue of surface instabilities, the inaccuracies here
are expected to manifest primarily in satellite galaxies moving through 
the intrahalo medium.  Such satellites should likely have more gas stripped 
than our EC-SPH algorithm predicts.  However, as we showed in
Figure~\ref{fig:mf_sat}, satellites are a strongly sub-dominant
contribution to the HIMF at all masses probed here.  Hence even if
two-phase instabilities should be causing all satellites to lose
their HI envelope, this will affect the HIMF by 25\% at
most.  A secondary
effect is that centrals partly build up their gas mass by accreting
gas-rich satellites.  However, \citet{ker05} and others have showed
that for star-forming galaxies, the contribution to the gas content
from infalling satellites (i.e. merging) is small compared to that
accreted directly from the IGM or in very small lumps.  This has
been shown for total gas, not necessarily \HI, and in any case
assessing the origin of \HI\ is tricky since it is a transient
reservoir rather than a steadily accumulating component such as
stars.  Hence we leave a detailed examination of the origin of \HI\
for the future, and preliminarily conclude that it seems unlikely
that the inaccuracies in EC-SPH are playing a major role in
establishing the \HI\ properties of galaxies, though this will
require more detailed testing with different codes to establish
conclusively.

\section{Summary and Discussion}\label{sec:summary}

The \HI\ content of galaxies offers a unique glimpse into baryon
cycling processes of inflow and outflow that are believed to govern
galaxy evolution.  Hydrodynamic simulations can provide a way to
connect \HI\ to such baryon cycling processes, as well as to help
interpret observations within a hierarchical galaxy formation
context.  In this work, we have examined the \HI\ properties of
galaxies in cosmological hydrodynamic simulations using \gad\ with
entropy-conserving SPH employing four different outflow prescriptions.
Our key conclusions can be summarised as follows: \begin{itemize}

\item The \HI\ mass function is strongly affected by the inclusion
of galactic outflows.  Without outflows, our simulations yield too
many galaxies at small \HI\ masses, and too few at large \HI\ masses.
Our model with a constant wind speed (cw) introduces a feature in
the HIMF around $M_{\rm HI}\sim 10^{10}M_\odot$ that is not seen
in the observations.  Our models with momentum-driven wind scalings
produce an HIMF that broadly agrees with the observations, particularly
at the low mass end down to our mass resolution limit of $\sim
10^8M_\odot$.  Switching to energy-driven scalings at low masses
($\sigma_{\rm gal}<75$~km/s) further improves the agreement, and
generates a low mass end HIMF slope of $-1.3$.  The HIMF is slightly
above the data at all masses, but likely within the systematic
uncertainties regarding how we compute \HI\ masses for our galaxies.
The low mass end slope becomes progressively steeper with redshift
out to $z\sim 3$, then remains constant at around $-2$ at higher
redshifts.  As an aside, we showed that the ezw model also provides
an excellent match, within the quoted uncertainties, to the galaxy
stellar mass function.

\item All models show that low-mass galaxies are more \HI\ rich,
relative to their stellar content, than high mass galaxies.  The
exact shape and amplitude of the relation depends on the
wind model.  All the models broadly agree with the GASS data that
probes down to $M_*\sim 10^{10}M_\odot$, but the different wind
models predict different trends to lower masses.  The constant wind
model predicts significantly lower gas richness than our vzw
(momentum-driven wind scalings) or ezw models at $M_*\la 10^{10}M_\odot$,
reflecting a strong suppression of inflow owing to energetic
winds~\citep{vdv10} that likewise suppresses the specific
SFRs~\citep{dav11a}.  Current observations favour a continuing
increase of gas richness to lower masses, and our ezw model again
fares slightly better than vzw at low masses, and both match much
better than the cw model.  There is essentially no redshift evolution
in the \HI\ richness as a function of stellar mass, showing that
the evolution of the HIMF is interconnected with the evolution of
the galaxy stellar mass function.

\item Galaxies with a high \HI\ content tend to have lower gas-phase
metallicities and higher star formation rates at a given $M_*$.
The deviation in sSFR and metallicity versus the deviation in \HI\
richness can be fit by power laws with slopes of $0.31$ and $-0.26$,
respectively.  The latter slope compares favourably with existing
observations of the \HI\ deficiency parameter versus metallicity
deviation.  These deviation trends can be understood within the
context of a model in which galaxies tend to have equilibrium
relations in star formation rate, metallicity, and gas content,
perturbed by stochastic accretion events that result in correlated
deviations in SFR, metallicity, star-forming gas mass, and \HI\
mass.  This suggests that higher \HI\ richness at a given $M_*$
is an indicator of recent accretion events in the outskirts of 
galaxies that stimulates star formation.

\item Environment plays an important role in governing \HI\ content,
causing an increasing suppression of the \HI\ content of satellites
in more massive halos.  The suppression is rapid, as indicated by
the bimodal distribution of the \HI\ fraction at a given halo mass.
Lower mass satellites are more likely to have their \HI\ removed.
The deviation of \HI\ with environment is anti-correlated with the
deviation in \HI\ richness, driven by \HI-poor satellite galaxies
in denser regions.

\item The global \HI\ mass density evolves slowly from $z\sim
5\rightarrow 0$, in broad agreement with the observations of DLAs
and other measures.  This is in contrast to the stellar mass density,
which grows substantially over that interval, highlighting the
transient nature of the \HI\ reservoir around galaxies.  Today, the
majority of the cosmic \HI\ mass is in galaxies with $10^9\la M_{\rm
HI}\la 10^{10}M_\odot$.  Our fiducial simulations predict $\Omega_{\rm HI}$ that
is lower than observed at $z\ga 2$, which could reflect either
the inability of these models to resolve the low-mass galaxies
that are a significant contribution to global \HI\ at high redshifts,
or else an evolving efficiency of converting \HI\ to $H_2$ owing
to generally lower metallicities at high-$z$.

\item We briefly examine numerical issues, including resolution convergence over a
modest span of a factor of 8 in mass resolution between two simulations
with momentum-driven winds.  The 
\HI\ mass function is well converged, but the \HI\ richness is not
converged at the $\sim 50\%$ level between the two simulations.  
Using a theoretically-based molecular gas criterion also results in 
up to $\sim 50\%$ difference in the HIMF at high redshifts.  There
may be further uncertainties owing to our use of entropy-conserving 
SPH, though it is expected to be small if the differences are
limited to satellite galaxies.  In general, it appears that our 
predictions are not free of numerical uncertainties, but seem to
be robust at the factor-of-two level, and hence the qualitative
conclusions presented above are generally robust.

\end{itemize}

Simultaneously matching the observed low mass end of the galactic
stellar mass function and the low mass end slope of the HIMF has
been challenging in simulations as well as in 
semi-analytic models (SAMs) of galaxy formation, where one has much
more freedom to adjust the physical model~\citep{mo05,lag11,lu12,lu13}.
Fundamentally, in most models this arises because the low mass end
slope of the dark matter halo mass function is quite steep, which
is exacerbated by the fact that low-mass galaxies are more \HI-rich.
\citet{mo05} argue that it is not possible to match the low mass
end of the HIMF in a very broad range of CDM based galaxy formation
models if one only makes two basic assumptions: first, that the
original gas distribution has the same specific angular momentum
as the dark matter halo, which it conserves as it forms an exponential
disk~\citep[as in][]{mo98}, and that the angular momentum distribution
of the gas and stars within the disk does not change subsequent to
its accretion onto the disk; second, that stars only form in gas
above a critical surface density $\Sigma_{crit}$, as observed
\citep{ken89}.  Many SAMs make these same assumptions.
\citet{mo05} populate some fraction of the cold dark matter halos
expected in a $\Lambda$CDM model with exponential gas disks and
assume that any region originally above $\Sigma_{crit}$ has its
surface density lowered to $\Sigma_{crit}$.  Making the (conservative)
assumption that half of the gas is neutral, they overproduce the
observed HIMF at the low mass end by a factor of more than five.
\citet{lu12} confirm this result by adjusting all their SAM parameters
using a Bayesian inference technique to match the observed low
redshift $K$-band galaxy luminosity function, and find that they
overpredict the observed HIMF by a large factor.

\citet{lu13} expand on this work, by adjusting their parameters to
simultaneously match both the observed low-$z$ $K$-band galaxy
luminosity function and the HIMF. They find that this is not possible
using standard models that include the above two assumptions, but
that it is possible if one does any one of the following: 1) reduce
$\Sigma_{crit}$ by about an order of magnitude, 2) allow the gas
disk to maintain an exponential profile while stars are forming,
i.e. allow gas from large radii beyond the star forming radius to
lose angular momentum by some process and move inwards, 3) include
some process that preheats the gas before it can be accreted into
a dark matter halo or have the formula by which one determines the
expected gas accretion rate be different for small mass halos. The
first option violates observations by a substantial margin, and
furthermore for smaller, lower-metallicity galaxies one observes
$\Sigma_{crit}$ to increase, not decrease~\citep{bol13}.  Our
simulations assume a three dimensional critical star formation
density of $n_{\rm H}=0.13 \cmc$, which approximately matches the
observed $\Sigma_{crit}$~\citep{spr03a}, and hence this is not
likely to be the difference between our models and the SAMs.  The
third possibility is also unlikely to be happening in our simulations,
since we do not explicitly add any sort of preheating, nor do we
explicitly vary the dark matter accretion rates from the usual
$\Lambda$CDM expectations~\citep[e.g.][]{nei08,fau11}. However, it
is possible that large scale structure formation itself creates its
own form of preheating \citep{mo05}.
Hence, if the arguments presented in \citet{lu13} apply to our
simulations, then since we simultaneously match the HIMF and the
stellar mass function, we suspect that some angular momentum transport
processes, either physical or numerical, causes the gas in our
simulated disks to move inwards.  

SAMs with different recipes than \citet{lu13} have had some success
matching the HIMF.  \citet{obr09}, \citet{pow10}, \citet{coo10} and 
\citet{lag11} all present various independent
SAMs that are broadly successful, though curiously they all tend 
to overpredict the observed HIMF by $\sim\times 3$ in the vicinity of 
$M_{\rm HI}\sim 10^{8-9} M_\odot$.  It is beyond the scope of this
paper to determine which ingredients in these SAMs differ from
the simple assumptions in \citet{lu12}, or whether the persistent
discrepancy in the dwarf mass range is a fundamental issue
or simply a matter of more parameter tuning.  Unfortunately, 
this illustrates a difficulty with SAMs that, owing to the many
plausible choices to achieve similar successes at matching data,
it can be difficult to extract robust interpretations for the 
key physical ingredients required.  This is particularly true
since the dynamics of the gas is crucial for modelling a transient
reservoir like \HI, but SAMs do not model the gas dynamics directly,
and must rely on assumptions to track this.  Nonetheless, SAMs offer the advantage 
of exploring a wide range of parameter space to better understand all
the various avenues by which models can reproduce observations.  Hence, SAMs and
hydrodynamical simulations provide complementary information about
how to build a successful model to match \HI\ and other properties
of galaxies and their associated gas.

With the emergence of numerous major new radio facilities such as
the JVLA, MeerKAT (Karoo Array Telescope), and ASKAP, and
eventually the SKA itself in the next decade, the future of \HI\
studies looks promising.  Simulations like the ones presented here,
hopefully improved substantially in the coming years, will be crucial
for providing a context to interpret these observations within the
broader multi-wavelength landscape of hierarchical galaxy evolution.  At this
point, it seems equally important both to probe the evolution of
the \HI\ content of galaxies and to go deeper around nearby galaxies
to better understand how \HI\ connects to the Cosmic Web.  Such
data promise to provide interesting new constraints on the key
processes of galaxy evolution such as inflows, outflows, and wind
recycling.  This paper represents a first step towards providing a
comprehensive framework for interpreting the wealth of forthcoming
\HI\ data.

 \section*{Acknowledgements}
The authors acknowledge helpful discussions with G. Kauffmann, C. Lagos,
Y. Lu, H. Mo, S. Rao, and the LADUMA team.
The simulations used here were run on the University of
Arizona's SGI cluster, ice, and on high-performance computing facilities 
owned by Carnegie Observatories.  This work was supported by the National
Science Foundation under grant numbers AST-0847667, AST-0907998, AST-0908334,
AST-0907651, and NASA grants NNX12AH86G and NNX10AJ95G.
RD acknowledges support from the South African Research Chairs Initiative 
and the South African National Research Foundation.
Computing resources were obtained through grant number DMS-
0619881 from the National Science Foundation and through a grant from
the Ahmanson Foundation.


\begin{thebibliography}{1}
\bibitem[Agertz et al. (2007)]{age07} Agertz, O. et al. 2007, MNRAS, 380, 963
\bibitem[Angles-Alcazar et al. (2013)]{ang13} Angl\'es-Alc\'azar, D., Dav\'e, R., \"Ozel, F., Oppenheimer, B. D. 2013, ApJ, submitted, arXiv:1303.6959
\bibitem[Andrews \& Martini(2013)]{and13} Andrews, B. \& Martini, P. 2013, ApJ, 765, 140
\bibitem[Asplund et al.(2009)]{asp09} Asplund, M., Grevesse, N., Sauval, A. J., Scott, P. 2009, ARA\&A, 47, 481
\bibitem[Baldry, Glazebrook, \& Driver(2008)]{bal08} Baldry, I. K., Glazebrook, K., Driver, S. P. 2008, MNRAS, 388, 945
\bibitem[Balsara et al.(1989)]{bal89} Balsara, D. S. 1989, Ph.D. Thesis, Univ. of Illinois, Champaign-Urbana
\bibitem[Battisti et al.(2012)]{bat12} Battisti, A. J. et al. 2012, ApJ, 744, 93
\bibitem[\protect\citeauthoryear{Becker et al.} {2012}]{bec12} Becker, G.D., Hewett, P.C., Worseck, G., Prochaska, J.X. 2012, MNRAS, submitted, arXiv:1208.2584
\bibitem[Bell et al.(2007)]{bel07} Bell, E. F., Zheng, X. Z., Papovich, C., Borch, A., Wolf, C., Meisenheimer, K. 2007, ApJ, 663, 834
\bibitem[Blanton et al.(2006)]{bla06} Blanton, M. R., Eisenstein, D., Hogg, D. W., Zehavi, I. 2006, ApJ, 645, 977
\bibitem[Blitz \& Rosolowsky(2006)]{bli06} Blitz, L. \& Rosolowsky, E. 2006, ApJ, 650, 933
\bibitem[Bolatto, Wolfire, \& Leroy(2013)]{bol13} Bolatto, A. D., Wolfire, M., Leroy, A. K. 2013, ARA\&A, 51, in press, arXiv:1301.3498 
\bibitem[\protect\citeauthoryear{Booth \& Schaye} {2009}]{boo09} Booth, C. M. \& Schaye, J. 2009, MNRAS, 398, 53
\bibitem[Bouch\'e et al.(2010)]{bou10} Bouche, N. et al. 2010, ApJ, 718, 1001
\bibitem[Braun(2012)]{bra12} Braun, R. 2012, ApJ, 749, 87
\bibitem[Brook et al.(2012)]{bro12} Brook, C. B., Stinson, G., Gibson, B. K., Roskar, R., Wadsley, J., Quinn, T. 2012, MNRAS, 419, 771
\bibitem[Catinella et al.(2010)]{cat10} Catinella, B. et al. 2010, MNRAS, 403, 683
\bibitem[Catinella et al.(2012)]{cat12} Catinella, B. et al. 2012, A\&A, 544, 65
\bibitem[\protect\citeauthoryear{Chabrier} {2003}]{cha03} Chabrier G., 2003, PASP, 115, 763
\bibitem[Cook et al.(2010)]{coo10} Cook, M., Evoli, C., Barausse, E., Granato, G. L., Lapi, A. 2010, MNRAS, 402, 941
\bibitem[Cortese et al.(2011)]{cor11} Cortese, L., Catinella, B., Boissier, S., Boselli, A., Heinis, S. 2011, MNRAS, 415, 1797
\bibitem[Daddi et al.(2007)]{dad07} Daddi, E. et al. 2007, ApJ, 670, 156
\bibitem[Delhaize et al.(2013)]{del13} Delhaize, J., Meyer, M., Stavely-Smith, L., Boyle, B. 2013, MNRAS, in press, arXiv:1305.1968
\bibitem[Dalla Vecchia \& Schaye(2008)]{dal08} Dalla Vecchia, C., Schaye, J. 2008, MNRAS, 387, 1431
\bibitem[Dav\'e(2008)]{dav08} Dav\'e, R. 2008, MNRAS, 385, 147
\bibitem[Dav\'e(2009)]{dav09} Dav\'e, R., ASPC, 419, 347
\bibitem[Dav\'e et al.(2010)]{dav10} Dav\'e, R., Oppenheimer, B. D., Katz, N., Kollmeier, J. A., Weinberg, D. H. 2010, MNRAS, 408, 2051
\bibitem[Dav\'e, Oppenheimer, \& Finlator(2011)]{dav11a} Dav\'e, R., Oppenheimer, B. D., Finlator, K. M. 2011, MNRAS, 415, 11
\bibitem[Dav\'e, Finlator, \& Oppenheimer(2011)]{dav11b} Dav\'e, R., Finlator, K. M., Oppenheimer, B. D. 2011, MNRAS, 416, 1354
\bibitem[Dav\'e, Finlator, \& Oppenheimer(2012)]{dav12} Dav\'e, R., Finlator, K. M., Oppenheimer, B. D. 2012, MNRAS, in press
\bibitem[Dekel \& Birnboim(2006)]{dek06} Dekel, A., Birnboim, Y. 2006, MNRAS, 368, 2
\bibitem[Dekel et al.(2009)]{dek09} Dekel, A. et al. 2009, Nature, 457, 451
\bibitem[Di Matteo et al.(2005)]{dim05} Di Matteo, T., Springel, V., Hernquist, L. 2005, Nature, 433, 604
\bibitem[Duffy et al.(2012)]{duf12} Duffy, A.R., Kay, S.T., Battye, R.A., Booth, C.M., Dalla Vecchia, C., Schaye, J. 2012, MNRAS, 420, 2799
\bibitem[Ellison et al.(2008)]{ell08} Ellison, S. L., Patton, D. R., Simard, L., McConnachie, A. W. 2008, ApJL, 672, L107
\bibitem[Erb et al.(2006)]{erb06} Erb, D. K., Shapley, A. E., Pettini, M., Steidel, C. C., Reddy, N. A., Adelberger, K. L. 2006, ApJ, 644, 813
\bibitem[Fardal et al. (2007)]{far07} Fardal, M. A., Katz, N., Weinberg, D. H., Dav\'e, R., Katz, N. 2007, MNRAS, 379, 985
\bibitem[Faucher-Giguere, Kere\v{s}, \& Ma(2010)]{fau10} Faucher-Giguere, C. A., Kere\v{s}, D., Dijkstra, M., Hernquist, L., Zaldarriaga, M. 2010, ApJ, 725, 633
\bibitem[Faucher-Giguere et al.(2011)]{fau11} Faucher-Giguere, C. A., Kere\v{s}, D., Ma, C.-P. 2011, 417, 2982
\bibitem[Finlator et al.(2006)]{fin06} Finlator, K., Dav\'e, R., Papovich, C., Hernquist, L. 2006, Apj, 639, 672
\bibitem[Finlator \& Dav\'e(2008)]{fin08} Finlator, K. \& Dav\'e, R. 2008, MNRAS, 385, 2181
\bibitem[Gabor et al.(2011)]{gab11} Gabor, J. M., Dav\'e, R., Oppenheimer, B. D., Finlator, K. M. 2011, MNRAS, 417, 2676
\bibitem[Gabor et al.(2012)]{gab12} Gabor, J. M. \& Dav\'e, R. 2012, MNRAS, 427, 1816
\bibitem[Geach et al.(2011)]{gea11} Geach, J. E., Smail, I., Moran, S. M., MacArthur, L. A., Lagos, C.d.P., Edge, A.C. 2011, ApJ, 730, 19
\bibitem[Giovanelli et al.(2005)]{gio05} Giovanelli, R. 2005, AJ, 130, 2598
\bibitem[Gnedin \& Kravtsov(2011)]{gne11} Gnedin, N. Y., Kravtsov, A. 2011, ApJ, 728, 88
\bibitem[Haardt \& Madau (2001)]{haa01} Haardt, F. \& Madau, P. 2001, in proc. XXXVIth Rencontres de Moriond, eds. D.M. Neumann \& J.T.T. Van.
\bibitem[Haynes et al.(1984)]{hay84} Haynes, M. P., Giovanelli, R., Chincarini, G. L. 1984, ARA\&A, 22, 445
\bibitem[Haynes et al.(2011)]{hay11} Haynes, M. P. et al. 2011, AJ, 142, 170
\bibitem[Hinshaw et al.(2009)]{hin09} Hinshaw, G. et al. 2009, ApJS, 180, 225
\bibitem[Holwerda et al.(2011)]{hol11} Holwerda, B. W. et al. 2012, in proc. IAU Symposium 284, ``The Spectral Energy Distribution of Galaxies" (SED2011), 5-9 Sep 2011, Preston, UK, eds. R.J. Tuffs \& C.C.Popescu
\bibitem[Hopkins \& Beacom(2006)]{hop06} Hopkins, A.M., Beacom, J.F. 2006, ApJ, 651, 142
\bibitem[Hopkins, Quataert, \& Murray(2012)]{hop12} Hopkins, P. F., Quataert, E., Murray, N. 2012, MNRAS, 412, 3522
\bibitem[Hopkins(2013)]{hop13} Hopkins, P. F. 2013, MNRAS, 428, 2840
\bibitem[Huang et al.(2012)]{hua12} Huang, S., Haynes, M. P., Giovanelli, R., Brinchmann, J., Stierwalt, S., Neff, S. G. 2012, AJ, 143, 133
\bibitem[Kassin et al.(2007)]{kas07} Kassin, S.A. et al. 2007, ApJ, 660, L35
\bibitem[Katz et al.(1996)]{kat96} Katz, N., Weinberg, D. H., Hernquist, L. 1996, ApJS, 105, 19
\bibitem[Kennicutt(1989)]{ken89} Kennicutt, R. C. 1989, ApJ, 344, 685
\bibitem[Kennicutt(1998)]{ken98} Kennicutt, R. C. 1998, ApJ, 498, 541
\bibitem[Kere\v{s} et al.(2005)]{ker05} Kere\v{s}, D., Katz, N., Weinberg, D. H., \& Dav\'e, R. 2005, MNRAS, 363, 2
\bibitem[Kere\v{s} et al.(2009)]{ker09} Kere\v{s}, D., Katz, N., Fardal, M., Dav\'e, R., Weinberg, D. H. 2009, MNRAS, 395, 160
\bibitem[Kirkman et al.(2007)]{kir07} Kirkman, D., Tytler, D., Lubin, D., Charloton, J. 2007, MNRAS, 376, 1227
\bibitem[Krumholz, McKee, \& Tumlinson(2008)]{kru08} Krumholz, M. R., McKee, C. F., Tumlinson, J. T. 2008, ApJ, 689, 865
\bibitem[Krumholz \& Gnedin(2011)]{kru11} Krumholz, M. R., Gnedin, N. Y. 2011, ApJ, 729, 36
\bibitem[Lagos et al.(2011)]{lag11} Lagos, C.d.P., Baugh, C. M., Lacey, C. G., Benson, A. J., Kim, H.-S., Power, C. 2011, MNRAS, 
\bibitem[Lara-L\'opez et al.(2010)]{lar10} Lara-L\'opez, M. A. et al. 2010, A\& 521, L53
\bibitem[Leitner(2012)]{lei12} Leitner, S. N. 2012, ApJ, 745, 149
\bibitem[Leroy et al.(2008)]{ler08} Leroy, A. K., Walter, F., Brinks, E., Bigiel, F., de Blok, W. J. G., Madore, B., Thornley, M. D. 2008, AJ, 136, 2782
\bibitem[Lu et al.(2012)]{lu12} Lu, Y., Mo, H. J., Katz, N., Weinberg, M. D. 2012, MNRAS, 421, 1779
\bibitem[Lu et al.(2013)]{lu13} Lu, Y., Mo, H. J., Katz, N., Weinberg, M. D. 2012, MNRAS, submitted
\bibitem[Mannucci et al.(2010)]{man10} Mannucci, F., Cresci, G., Maiolino, R., Marconi, A., Gnerucci, A. 2010, MNRAS, submitted, arXiv:1005.0006
\bibitem[McGaugh et al.(2010)]{mcg10} McGaugh, S.S., Schombert, J.M., de Blok, W.J.G., Zagursky, M.J. 2010, ApJ, 708, L14
\bibitem[McGaugh et al.(2012)]{mcg12} McGaugh, S.S. 2012, AJ, 143, 40
\bibitem[\protect\citeauthoryear{McKee \& Ostriker} {1977}]{mck77} McKee, C. F. \& Ostriker, J. P.\ 1977, ApJ, 218, 148
\bibitem[Meyer et al.(2004)]{mey04} Meyer, M. J. et al. 2004, MNRAS, 350, 1195
\bibitem[Mo, Mao, \& White(1998)]{mo98} Mo, H.J., Mao, S., White, S.D.M. 1998, MNRAS, 295, 319
\bibitem[Mo et al (2005)]{mo05} Mo, H. J., Yang, X., van den Bosch, F.C., \& Katz, N. 2005, MNRAS, 363, 1155
\bibitem[Moldar et al.(2009)]{mol09} Moldar, S.M., Hearn, N., Haiman, Z., Bryan, G., Evrad, A.E., Lake, G. 2009, ApJ, 696, 1640
\bibitem[Moran et al.(2012)]{mor12} Moran, S.M. et al. 2012, ApJ, 745, 66
\bibitem[Murray, Quataert, \& Thompson(2005)]{mur05} Murray, N., Quatert, E., Thompson, T. A. 2005, ApJ, 618, 569
\bibitem[Murray, Quataert, \& Thompson(2010)]{mur10} Murray, N., Quatert, E., Thompson, T. A. 2010, ApJ, 709, 191
\bibitem[Narayanan, Bothwell, \& Dav\'e(2012)]{nar12} Narayanan, D., Bothwell, M., Dav\'e, R. 2012, MNRAS, 426, 1178
\bibitem[Neistein \& Dekel (2008)]{nei08} Neistein, E., Dekel, A. 2008, MNRAS, 388, 1792
\bibitem[Noterdaeme et al.(2012)]{not12} Noterdaeme, P. et al. 2012, A\&A, 547, L1
\bibitem[Obreschkow et al.(2009)]{obr09} Obreschkow, D., Croton, D., De Lucia, G., Khochfar, S., Rawlings, S. 2009, ApJ, 698, 1467
\bibitem[Oppenheimer et al.(2010)]{opp10} Oppenheimer, B.D., Dav\'e, R., Kere\v{s}, D., Katz, N., Kollmeier, J.A., Weinberg, D.H. 2010, MNRAS, 406, 2325
\bibitem[Oppenheimer \& Dav\'e(2006)]{opp06} Oppenheimer, B. D. \& Dav\'e, R. 2006, MNRAS, 373, 1265
\bibitem[Oppenheimer \& Dav\'e(2008)]{opp08} Oppenheimer, B. D. \& Dav\'e, R. 2008, MNRAS, 387, 577
\bibitem[Popping et al.(2009)]{pop09} Popping, A., Dav\'e, R., Braun, R., Oppenheimer, B. D. 2009, A\&A, 504, 15
\bibitem[Power et al.(2010)]{pow10} Power C., Baugh C.M., Lacey C.G., 2010, MNRAS, 406, 43
\bibitem[Prochaska \& Wolfe(2009)]{pro09} Prochaska, J.X., Wolfe, A.M. 2009, ApJ, 696, 1543
\bibitem[Rao et al.(2006)]{rao06} Rao, S.M., Turnshek, D.A., Nestor, D.B. 2006, ApJ, 636, 610
\bibitem[Read \& Hayfield(2012)]{rea12} Read, J. I., Hayfield, T. 2012, MNRAS, 422, 3037
\bibitem[Richard et al.(2011)]{ric11} Richard, J., Jones, T., Ellis, R., Stark, D. P., Livermore, R., Swinbank, M. 2011, MNRAS, in press
\bibitem[Robertson, Shields, \& Blanc(2012)]{rob12} Robertson, P., Shields, G.A., Blanc, G.A. 2012, ApJ, 748, 48
\bibitem[Rupke, Veilleux \& Sanders(2005)]{rup05} Rupke, D. S., Veilleux, S., \& Sanders, D. B. 2005, ApJS, 160, 115
\bibitem[Saitoh \& Makino(2013)]{sai13} Saitoh, T. R., Makino, J. 2013, ApJ, 768, 44
\bibitem[Salim et al.(2007)]{sal07} Salim, S. 2007, ApJS, 173, 267
\bibitem[\protect\citeauthoryear{Schmidt} {1959}]{sch59} Schmidt, M.\ 1959, ApJ, 129, 243
\bibitem[Springel \& Hernquist(2003)]{spr03a} Springel, V. \& Hernquist, L. 2003, MNRAS, 339, 289
\bibitem[Springel \& Hernquist(2003b)]{spr03b} Springel, V. \& Hernquist, L. 2003, MNRAS, 339, 312
\bibitem[Springel(2005)]{spr05} Springel, V. 2005, MNRAS, 364, 1105
\bibitem[Tacconi et al.(2010)]{tac10} Tacconi, L. J. et al. 2010, Nature, 463, 781 
\bibitem[Tacconi et al.(2012)]{tac12} Tacconi, L. J. et al. 2012, ApJ, submitted, arXiv:1211.5743
\bibitem[Torrey et al.(2012)]{tor12} Torrey, P., Vogelsberger, M., Sijacki, D., Springel, V., Hernquist, L. 2012, MNRAS, 427, 2224
\bibitem[Tremonti et al.(2004)]{tre04} Tremonti, C. A. et al. 2004, ApJ, 613, 898
\bibitem[Tully \& Fisher(1977)]{tul77} Tully, R.B., Fisher, J.R. 1977, A\&A, 54, 661
\bibitem[van de Voort et al.(2010)]{vdv10} van de Voort, F., Schaye, J., Booth, C. M., Haas, M. R., Dalla Vecchia, C. 2010, MNRAS, submitted, arXiv:1011.2491
\bibitem[Verner \& Ferland(1996)]{ver96} Verner, D. A. \& Ferland, G. J. 1996, ApJS, 103, 467
\bibitem[Vogt et al.(2004)]{vog04} Vogt, N.P., Haynes, M.P., Herter, T., Giovanelli, R. 2004, AJ, 127, 3273
\bibitem[Weinmann et al.(2012)]{wei12} Weinmann, S. M., Pasquali, A., Oppenheimer, B. D., Finlator, K., Mendel, J. T., Crain, R. A., Maccio, A. V. 2012, MNRAS, accepted
\bibitem[Wiersma et al.(2009)]{wie09} Wiersma, R.P., Schaye, K., Theuns, T., Dalla Vecchia, C., Tornatore, L. 2009, 399, 574
\bibitem[Zwaan et al.(2005)]{zwa05} Zwaan, M. A., Meyer, M. J., Staveley-Smith, L., Webster, R. L. 2005, MNRAS, 359, L30


\end{thebibliography}
\end{document}